\documentclass[pra,aps,onecolumn,nopacs,nofootinbib]{revtex4}
\usepackage{graphicx}  \usepackage{dcolumn}  \usepackage{bm} \usepackage{amsmath}  \usepackage{amsfonts}  \usepackage{epsfig} \usepackage{fancyhdr}  \usepackage{enumerate}  \usepackage{float} \usepackage{setspace}  \usepackage{fancybox}

\def\bra#1{\mathinner{\langle{#1}|}}
\def\ket#1{\mathinner{|{#1}\rangle}}

\def\Bra#1{\left<1>}
{\catcode`\|=\active\gdef\Braket#1{\left<\mathcode`\|"8000\let|\bravert {#1}\right>}}
\def\bravert{\egroup\,\vrule\,\bgroup}

\newcommand{\ee}{\textbf{e}}  \newcommand{\kk}{\textbf{k}}
\newcommand{\kkg}{\textbf{k}} \newcommand{\rr}{\textbf{r}}
\newcommand{\RR}{\textbf{R}}  \newcommand{\zz}{\textbf{z}}

\newcommand{\ag}{\hat{a}}          \newcommand{\QQ}{\textbf{Q}}
\newcommand{\HH}{\hat{\textbf{H}}} \newcommand{\HHs}{\textbf{H}}
\newcommand{\EE}{\hat{\textbf{E}}} \newcommand{\EEs}{\textbf{E}}
\newcommand{\pp}{\textbf{p}}       \newcommand{\mm}{\textbf{m}}
\newcommand{\ggg}{\mathcal{G}}     \newcommand{\II}{\mathbb{I}_3}
\newcommand{\al}{\alpha}           

\newcommand{\A}{A(d)}     \newcommand{\B}{B(d)}
\newcommand{\C}{C(d)}
\newcommand{\Ad}{A}     \newcommand{\Ada}{A^*}
\newcommand{\Bd}{B}     \newcommand{\Cd}{C}
\newcommand{\Cda}{C^*}  \newcommand{\Add}{A^2}
\newcommand{\Addd}{A^3} \newcommand{\Addda}{\left(A^*\right)^{3}}
\newcommand{\Bdd}{B^2}  \newcommand{\Cdd}{C^2}
\newcommand{\Cddd}{C^3} \newcommand{\Cddda}{\left(C^{*}\right)^3}

\newcommand{\I}{{\rm Im}}  \newcommand{\R}{{\rm Re}}
\newcommand{\w}{\omega}
\newcommand{\GG}{\mathcal{G}}

\newcommand{\un}{u_{\nu}}  \newcommand{\unp}{u_{\nu'}}
    
\newcommand{\wn}{w_{\nu}}  \newcommand{\wnp}{w_{\nu'}}
    
\newcommand{\tn}{t_{\nu}}  \newcommand{\dn}{g_{\nu}}

\newcommand{\aep}{\alpha_{1}^{\rm E}}      \newcommand{\amp}{\alpha_{1}^{\rm M}}
\newcommand{\aes}{\alpha_{2}^{\rm E}}      \newcommand{\ams}{\alpha_{2}^{\rm M}}
\newcommand{\apn}{\alpha_{1}^{\rm \nu}}    \newcommand{\asn}{\alpha_{2}^{\rm \nu}}
\newcommand{\apnp}{\alpha_{1}^{\rm \nu'}}  \newcommand{\asnp}{\alpha_{2}^{\rm \nu'}}

\newcommand{\chipn}{\chi_{1}^{\rm \nu}}   \newcommand{\chisn}{\chi_{2}^{\rm \nu}}

\def\qh{\hat{q}}  \def\Hh{\hat{H}}
\def\ph{\hat{p}}

\begin{document}
\title{Radiative Heat Transfer between Neighboring Particles}
\author{Alejandro Manjavacas}
\email{a.manjavacas@csic.es}
\affiliation{IQFR - CSIC, Serrano 119, 28006 Madrid, Spain}
\author{F. Javier Garc\'{\i}a de Abajo}
\email{J.G.deAbajo@csic.es}
\affiliation{IQFR - CSIC, Serrano 119, 28006 Madrid, Spain}

\date{\today}

\begin{abstract}
The near-field interaction between two neighboring particles is known to produce enhanced radiative heat transfer. We advance in the understanding of this phenomenon by including the full electromagnetic particle response, heat exchange with the environment, and important radiative corrections both in the distance dependence of the fields and in the particle absorption coefficients. We find that crossed terms of electric and magnetic interactions dominate the transfer rate between gold and SiC particles, whereas radiative corrections reduce it by several orders of magnitude even at small separations. Radiation away from the dimer can be strongly suppressed or enhanced at low and high temperatures, respectively. These effects must be taken into account for an accurate description of radiative heat transfer in nanostructured environments.
\end{abstract}

\maketitle \tableofcontents

\section{Introduction}

Blackbody radiation mediates heat exchange between bodies placed in vacuum and separated by large distances $d$ compared to the thermal wavelength $\lambda_T=2\pi\hbar c/k_BT$. For parallel plates, this leads to a radiative heat transfer (RHT) rate independent of $d$. However, when $d\ll\lambda_T$, the rate is enhanced by several orders of magnitude due to the involvement of evanescent waves. Pioneering measurements \cite{H1969,DBT1970} revealed this phenomenon, which was first explained in terms of near-field fluctuations \cite{PV1971}. After a long series of experimental \cite{NSH09,OQW11} and theoretical \cite{PV1971,LM94,CG99,SJC00,VP01,NC03,VP04,VP07,M08,BZF09} studies, recent observations have accurately confirmed a $1/d^2$ dependence for sapphire plates at room temperature down to $d\sim1\,\mu$m \cite{OQW11}, and also a $1/d$ dependence for large silica spheres placed near a silica plate down to $d\sim30\,$nm \cite{RSJ09}, although these laws can be substantially corrected by nonlocal  \cite{VP04}, phonon \cite{MJC01,PM10}, and photonic crystal \cite{RIB11} effects. In this context, the interaction of a particle with a plate has been explored both from experimental \cite{KMP05,NSH09,RSJ09,SNC09} and theoretical \cite{P99,MJC01,VP07,CLV08,KEK11,OF11} fronts. Modeling heat exchange between two \cite{DVJ05,NC08,CLV08_1,PRL08,PLR09,DK10,MA11} or more \cite{BBJ11} particles has been the subject of intense activity as well.

Magnetic polarization has been claimed to dominate RHT between metallic nanoparticles \cite{CLV08_1}. However, electromagnetic crossed terms (EMCTs, i.e., terms mixing the electric and magnetic particle responses) have been ignored so far, although they could play a leading role in transfers within heterogeneous structures. Likewise, radiative corrections in the absorption of dielectric particles deserve further consideration, as we show below. Thus, the current level of understanding of RHT between two particles appears to be incomplete.

Here, we formulate a complete solution of RHT between two nanoparticles within the assumption of dipolar response. We show that EMCTs are dominant in combinations of metallic and dielectric particles, such as gold and SiC. We introduce a relevant retardation correction beyond the customary treatment of  polarization fluctuations, which results in a sizable reduction in the predicted transfer rate. Furthermore, we show that heat losses into the environment can be either dominant or negligible depending on the temperature and particle composition. An accurate description of RHT in nanostrutured environments requires incorporating these effects, for which the two-particle system discussed here provides a tutorial approach as well as an estimate of the importance of EMCTs, radiative corrections, and interaction with the environment.

\begin{figure}[b]
\begin{center}
\includegraphics[width=65mm,angle=0,clip]{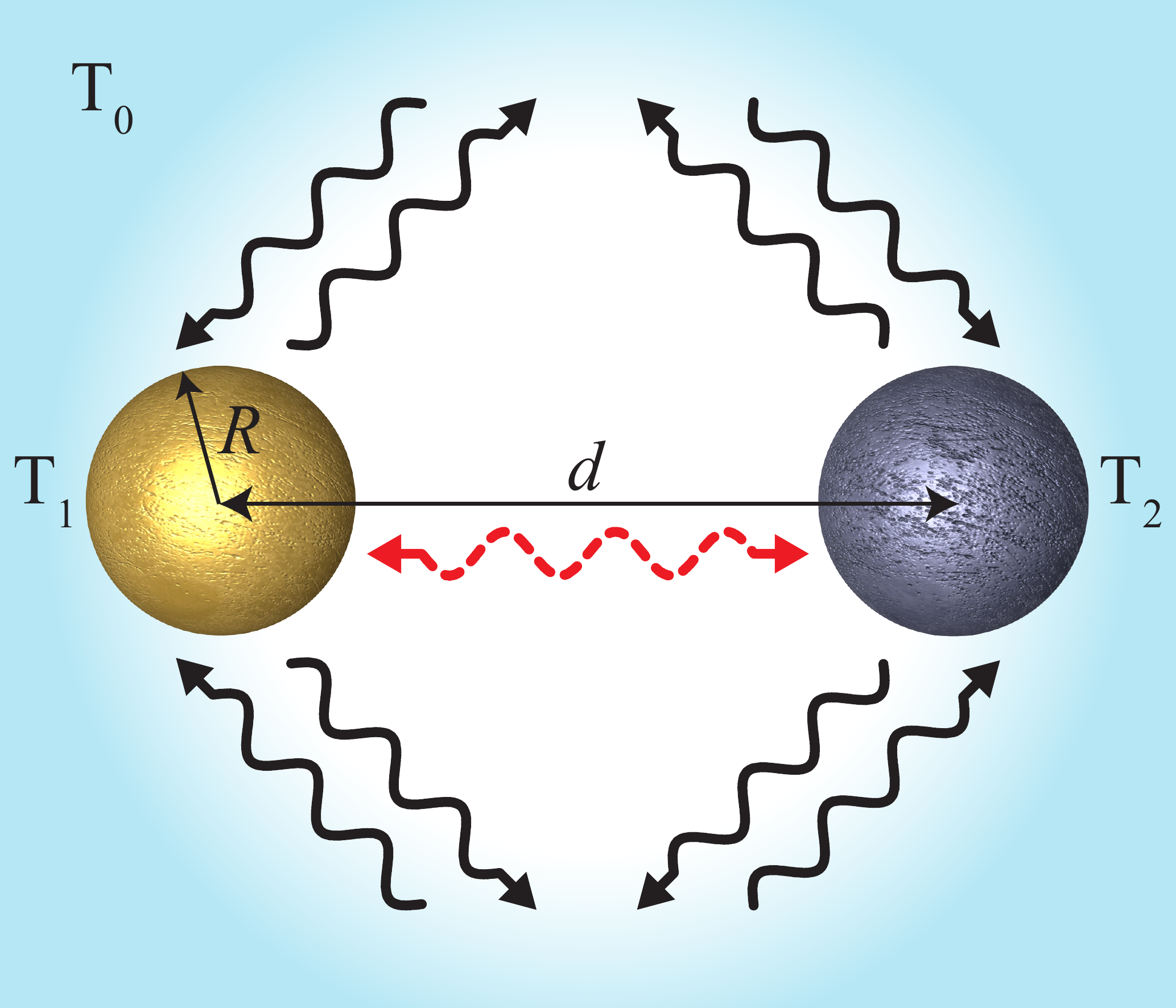}
\caption{(color online). Description of the system under study. Two particles of radius $R$ at temperatures $T_1$ and $T_2$ separated by a distance $d$ are placed in a vacuum at temperature $T_0$. Each particle exchanges thermal energy with the other particle and with the surrounding vacuum.} \label{Fig1}
\end{center}
\end{figure}

\section{Description of the model}

We consider two spherical particles of radius $R$ at temperatures $T_1$ and $T_2$ separated by a center-to-center distance $d$ along the $z$ direction and placed in a vacuum at temperature $T_0$, as shown in Fig.\ \ref{Fig1}. We focus on small particles such that $R\ll d,\,\lambda_T$, so that their responses can be described through the polarizabilities $\alpha_1$ and $\alpha_2$ \cite{NC08} (see Appendix, Fig.\ 6). RHT between the particles and the environment is produced by fluctuations in the vacuum electromagnetic field and the particle dipoles. We simplify the notation by combining electric and magnetic field components acting on each particle $j=1,2$, as well as electric ($p$) and magnetic ($m$) dipoles, in the vectors
\begin{equation}
E_j=\left( \begin{array}{c}
E_{j,x} \\ E_{j,y} \\ E_{j,z} \\
H_{j,x} \\ H_{j,y} \\ H_{j,z}
\end{array} \right),
\ \ \ \ p_j=\left( \begin{array}{c}
p_{j,x} \\ p_{j,y} \\ p_{j,z} \\
m_{j,x} \\ m_{j,y} \\ m_{j,z}
\end{array} \right), \nonumber
\end{equation}
respectively. Likewise, the polarizability tensor becomes
\begin{equation}
\alpha_j=\left( \begin{array}{cc}
\alpha_j^E\mathbb{I}_3 & 0 \\
0  & \alpha_j^M\mathbb{I}_3 \\
\end{array} \right),\nonumber
\end{equation}
where $\mathbb{I}_3$ is the $3\times3$ identity matrix and the $E$ ($M$) subscript refers to electric (magnetic) components.

The net power absorbed by particle $1$ is the sum of dipole and field fluctuation terms,
\begin{equation}\label{18}
\mathcal{P}_1=\mathcal{P}_1^{\rm field}+\mathcal{P}_1^{\rm dip}.\nonumber
\end{equation}
More precisely (see Appendix),
\begin{align}
P_1^{\rm field}=&\int_{-\infty}^{\infty}\frac{d\omega d\omega'}{(2\pi)^2} e^{-i(\omega-\omega')t}\,\omega' \label{int1} \\
&\times\left\langle E_1^{+}(\omega')
\left[i\alpha^+(\omega)-(2k^3/3)|\alpha(\omega)|^2\right]
E_1(\omega)\right\rangle \nonumber
\end{align}
represents the work exerted by the fluctuating field on particle 1. Here, $k=\omega/c$ is the wave vector of light at frequency $\omega$. Moreover, the self-consistent fields $E_j$ include the response of the system to the fluctuating source fields $E_j^{\rm fl}$ via the relations
\begin{align}
E_1&=E_1^{\rm fl}+\GG_{12}\alpha_2E_2, \nonumber\\
E_2&=E_2^{\rm fl}+\GG_{21}\alpha_1E_1, \nonumber
\end{align}
where $\GG_{12}$ is the distance-dependent dipole-dipole inter-particle interaction,
\begin{equation}
\GG_{12}=\left( \begin{array}{cccccc}
A &  0 &  0 &  0 & -C &  0 \\
 0 & A &  0 & C &  0 &  0 \\
 0 &  0 & B &  0 &  0 &  0 \\
 0 & C &  0 & A &  0 &  0 \\
-C&  0 &  0 &  0 & A &  0 \\
 0 &  0 &  0 &  0 &  0 & B \\
\end{array} \right),\nonumber
\end{equation}
%\begin{align}
%A&=\frac{e^{ikd}}{d^3}[(kd)^2+ikd-1], \nonumber\\
%B&=\frac{e^{ikd}}{d^3}2(1-ikd), \nonumber\\
%C&=\frac{e^{ikd}}{d^3}[(kd)^2+ikd], \nonumber
%\end{align}
\begin{align}
A&=\exp(ikd)\,(k^2/d+ik/d^2-1/d^3), \nonumber\\
B&=\exp(ikd)\,2(1/d^3-ik/d^2), \nonumber\\
C&=\exp(ikd)\,(k^2/d+ik/d^2), \nonumber
\end{align}
and $\GG_{21}$ takes the same form as $\GG_{12}$ with $C$ replaced by $-C$. In Eq.\ (\ref{int1}), $\langle\rangle$ represents the average over field fluctuations, which we perform by applying the fluctuation-dissipation theorem (FDT) \cite{R1959_2,paper157} (see Appendix)
\begin{align}
\left\langle \left[E_j^{\rm fl}(\omega)\right]^+E_{j'}^{\rm fl}(\omega') \right\rangle =4\pi \hbar\delta(\omega-\omega') \eta_{jj'}(\omega)\left[n_0(\omega)+\frac{1}{2}\right], \nonumber
\end{align}
where $\eta_{12}=(1/2)\left[\I\{\GG_{12}+\GG_{21}\}+i \R\{\GG_{12}-\GG_{21}\}\right]$,
$\eta_{21}=\eta_{12}^{+}$, $\eta_{11}=\eta_{22}=(2k^3/3)\mathbb{I}_6$, and $n_0(\omega)=[\exp(\hbar\omega/k_BT_0)-1]^{-1}$ is the Bose-Einstein distribution at the vacuum temperature $T_0$.

Likewise, the contribution of fluctuating dipoles is (see Appendix)
\begin{align}
P_1^{\rm dip}&=\int_{-\infty}^{\infty}\frac{d\omega d\omega'}{(2\pi)^2} e^{-i(\omega-\omega')t} \label{int2} \\
&\big[i\omega'\left\langle p_1^{+}(\omega')\GG_{12}(\omega)p_2(\omega)\right\rangle-\frac{2\omega^4}{3c^3} \left\langle p_1^{+}(\omega')p_1(\omega)\right\rangle\big],\nonumber
\end{align}
where the first term inside the square brackets accounts for the effect of the field produced by particle 2 on particle 1, while the second term describes the interaction of the particle dipole with the vacuum. The self-consistent dipoles satisfy the relations
\begin{align}
p_1&=p_1^{\rm fl}+\alpha_1\GG_{12}p_2,\nonumber\\
p_2&=p_2^{\rm fl}+\alpha_2\GG_{21}p_1,\nonumber
\end{align}
where $p_j^{\rm fl}$ is the fluctuating source dipole at particle $j$. The relevant FDT now becomes \cite{R1959_2,paper157} (see Appendix)
\begin{align}
\left\langle\left[p_{j}^{\rm fl}(\omega)\right]^+p_{j'}^{\rm fl}(\omega') \right\rangle =4\pi \hbar \delta(\omega-\omega')\delta_{jj'} \chi_j(\omega)\left[n_j(\omega)+\frac{1}{2}\right], \nonumber
\end{align}
where we use
\begin{equation}
\chi_j=\I\{\alpha_j\}-(2k^3/3)|\alpha_j|^2\mathbb{I}_6, \label{chi}
\end{equation}
rather than $\I\{\alpha_j\}$ in order to prevent non-absorbing particles from undergoing unphysical fluctuations. We set the polarizabiltiy to $\alpha_j^\nu=(3/2k^3)t_{j,1}^\nu$ ($\nu=E,M$), where $t_{i,1}^\nu$ is the dipole Mie scattering coefficient (see  Appendix). This definition of $\alpha_j$ complies with the optical theorem condition \cite{V1981} $\I\{\alpha_j\}\ge(2k^3/3)|\alpha_j|^2$, where the equality applies to non-absorbing particles ($\chi_j=0$). Incidentally, dipole and field fluctuations originate in different physical systems, and therefore, there are not crossed terms between the two of them.

Finally, using the FDT to evaluate the integrals of Eqs.\ (\ref{int1}) and (\ref{int2}), we find, after some lengthy but straightforward algebra (see  Appendix),
\begin{align}
\mathcal{P}_{1}=\frac{2\hbar}{\pi}\sum_{\nu={E,M}}\int_{0}^{\infty}\omega d\omega\;\chi_1^\nu \sum_{i=1}^2\left(\frac{2\Gamma_{i\perp}^\nu}{|s|^2}+\frac{\Gamma_{i\parallel}^\nu}{|t_\nu|^2} \right),\label{Pall}
\end{align}
where
\begin{align}
&\Gamma_{1\perp}^\nu=\bigg[\frac{2k^3}{3} \left(\left|u_\nu\right|^2+\left|w_\nu\right|^2\right)\nonumber\\
& \ \ \ \ +\I\big\{\alpha_2^\nu\left[A u_\nu-g_\nu C w_\nu\right]\left[A u_\nu^{*}+g_\nu C w_\nu^{*}\right]\big\}\nonumber\\
& \ \ \ \ +\I\big\{\alpha_2^{\nu'}\left[A w_\nu-g_\nu C u_\nu\right]\left[A w_\nu^{*}+g_\nu C u_\nu^{*}\right]\big\}\bigg](n_0-n_1),\nonumber\\
&\Gamma_{2\perp}^\nu=\big(\chi_2^\nu\left|A u_\nu-g_\nu C w_\nu\right|^2 +\chi_2^{\nu'}\left|A w_\nu-g_\nu C u_\nu\right|^2\big)\;(n_2-n_0), \nonumber \\
&\Gamma_{1\parallel}^\nu=\left[\frac{2k^3}{3}+
\I\left\{\alpha_2^\nu B^2\right\}\right](n_0-n_1), \nonumber\\
&\Gamma_{2\parallel}^\nu=\chi_2^\nu\left|B\right|^2\;(n_2-n_0), \nonumber\\
&u_\nu=1-\alpha_{1}^{\nu'}\alpha_{2}^{\nu'}A^2+\alpha_{1}^{\nu'}\alpha_{2}^\nu C^2, \nonumber\\
&w_\nu=\alpha_{1}^{\nu'}\left(\alpha_{2}^E-\alpha_{2}^M\right)AC, \nonumber\\
&{\rm s}=1-\alpha_1^E\alpha_2^E A^2-\alpha_1^M\alpha_2^M A^2+\alpha_1^E\alpha_2^M C^2+\alpha_1^M\alpha_2^E C^2\nonumber\\ &\;\;\;\;\;\;+\alpha_1^E\alpha_2^E\alpha_1^M\alpha_2^M\left(A^2-C^2\right)^2, \nonumber\\
&{\rm t}_\nu=1-\alpha_{1}^\nu\alpha_{2}^\nu B^2, \nonumber
\end{align}
and $\nu'=M$ ($E$) and $g_\nu=+1$ ($-1$) when $\nu=E$ ($M$).

Neglecting the magnetic response ($\alpha_j^M=0$, $C=0$), multiple scattering ($u_\nu=t_\nu=s=1$, $w_\nu=0$), and radiative corrections in the particles response ($\chi_j=\I\{\alpha_j\}$), the above expressions reduce to
\begin{align}
&\mathcal{P}_1=\frac{4\hbar}{\pi}\int_{0}^{\infty}\omega\,d\omega\; \I\left\{\alpha_1^E\right\} \; \bigg\{\Big[k^3
+2\R\left\{\alpha_2^EA\right\}\I\{A\} \nonumber\\
& \ \ \ \ \ \ \ \ +\R\left\{\alpha_2^EB\right\}\I\{B\}\Big] (n_0-n_1)\nonumber\\
& \ \ \ \ +\I\left\{\alpha_2^E\right\}\left(|A|^2+|B|^2/2\right)\;(n_2-n_1)\bigg\}, \nonumber
\end{align}
where the $n_2-n_1$ term describes direct RHT between the dimer particles and coincides with a previously reported expression \cite{DVJ05,CLV08,DK10}. The remaining $n_0-n_1$ term accounts for heat exchange between particle 1 and the surrounded vacuum, partially assisted by the presence of particle 2.

\begin{figure}
\begin{center}
\includegraphics[width=85mm,angle=0,clip]{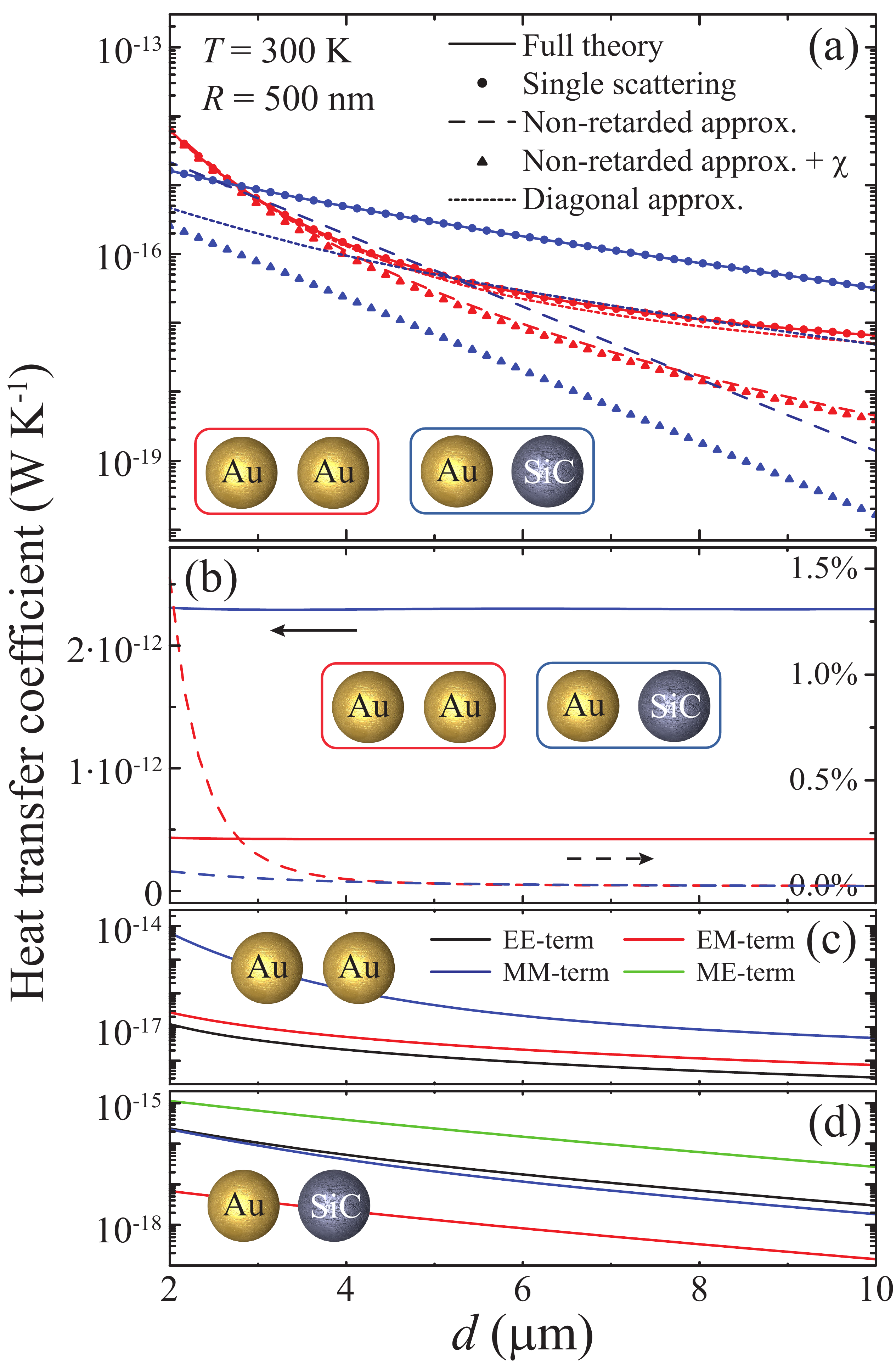}
\caption{(color online). Dependence of the radiative heat transfer coefficient (HTC) on particle separation $d$. The HTC from particle 2 (right) to particle 1 (left) is defined as $\mathcal{P}_1/\delta T$ with $T_1=T_0=T$ and $T_2=T+\delta T$ (see Fig.\ \ref{Fig1}). {\bf (a)} HTC at $T=300$\,K obtained from our full theory (solid curves), compared to the result of neglecting multiple scattering (circles), retardation effects everywhere (dashed curves), retardation effects except in $\chi_j$ (triangles), or EMCTs (dotted curves). Gold-gold (red curves) and gold-SiC (blue curves) dimers are considered. The particles radius is $R=500$\,nm. {\bf (b)} Heat power lost by particle 2  (solid curves) and fraction of that power absorbed by particle 1 (broken curves, right scale). {\bf (c,d)} Electric-electric (black), electric-magnetic (red), magnetic-magnetic (blue), and magnetic-electric (green) partial contributions to the HTC in the gold-gold (c) and gold-SiC (d) dimers.} \label{Fig2}
\end{center}
\end{figure}

\section{Results and discussion}

\begin{figure}
\begin{center}
\includegraphics[width=85mm,angle=0,clip]{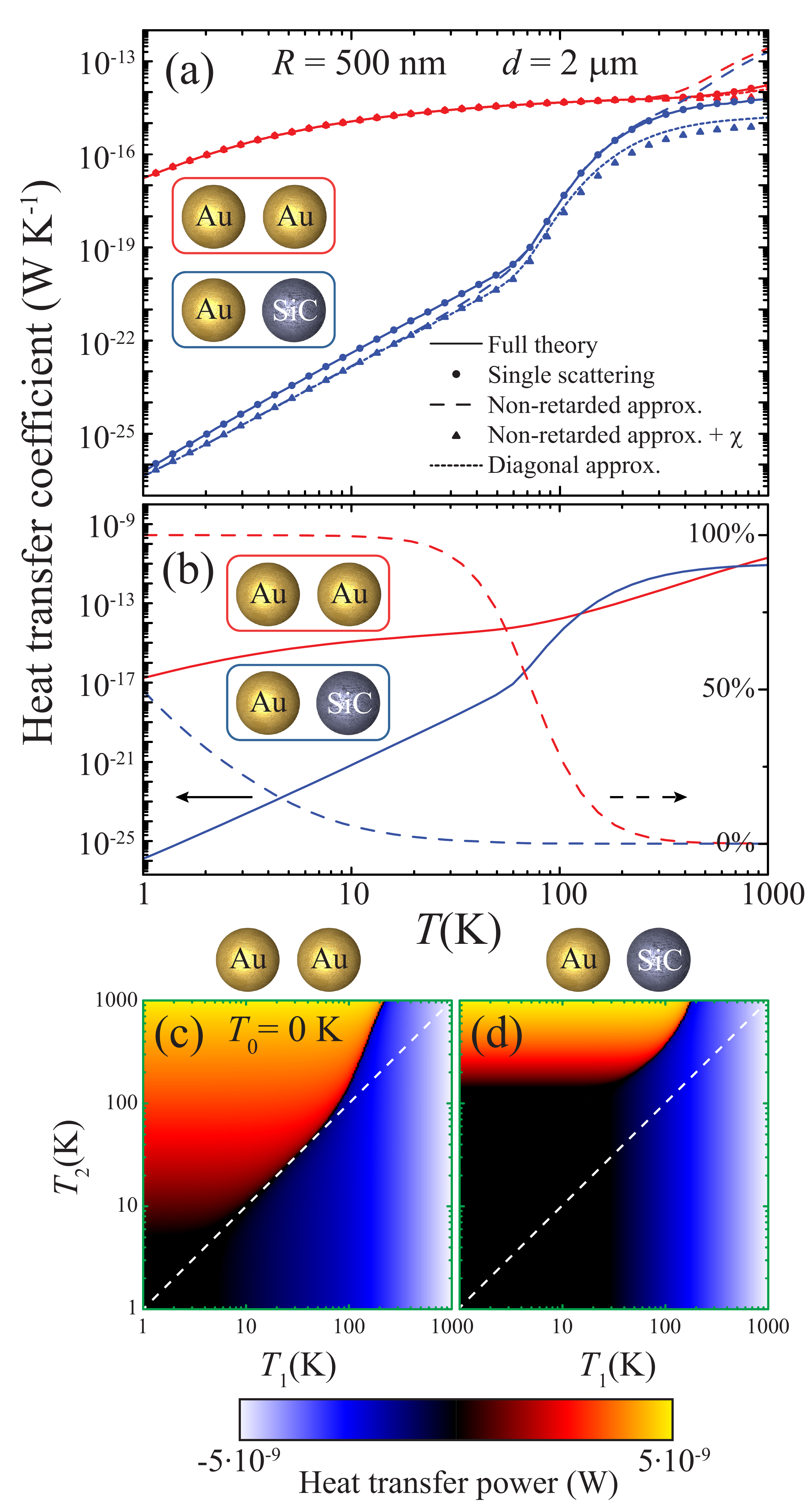}
\caption{(color online). Temperature dependence of RHT. {\bf (a)} HTC for the particles of Fig.\ \ref{Fig2}. {\bf (b)} Heat power lost by the right particle and fraction absorbed by the left particle. {\bf (c)} Power absorbed (red-yellow scale, positive) or emitted (blue-white scale, negative) by the left particle of a gold-gold dimer as a function of $T_1$ and $T_2$ with the vacuum at $T_0=0$ (see Fig.\ \ref{Fig1}). {\bf (d)} Same as (c) for a gold-SiC dimer. The particle distance is $d=2\,\mu$m and the radius is $R=500\,$nm in all cases.} \label{Fig3}
\end{center}
\end{figure}

\begin{figure}
\begin{center}
\includegraphics[width=85mm,angle=0,clip]{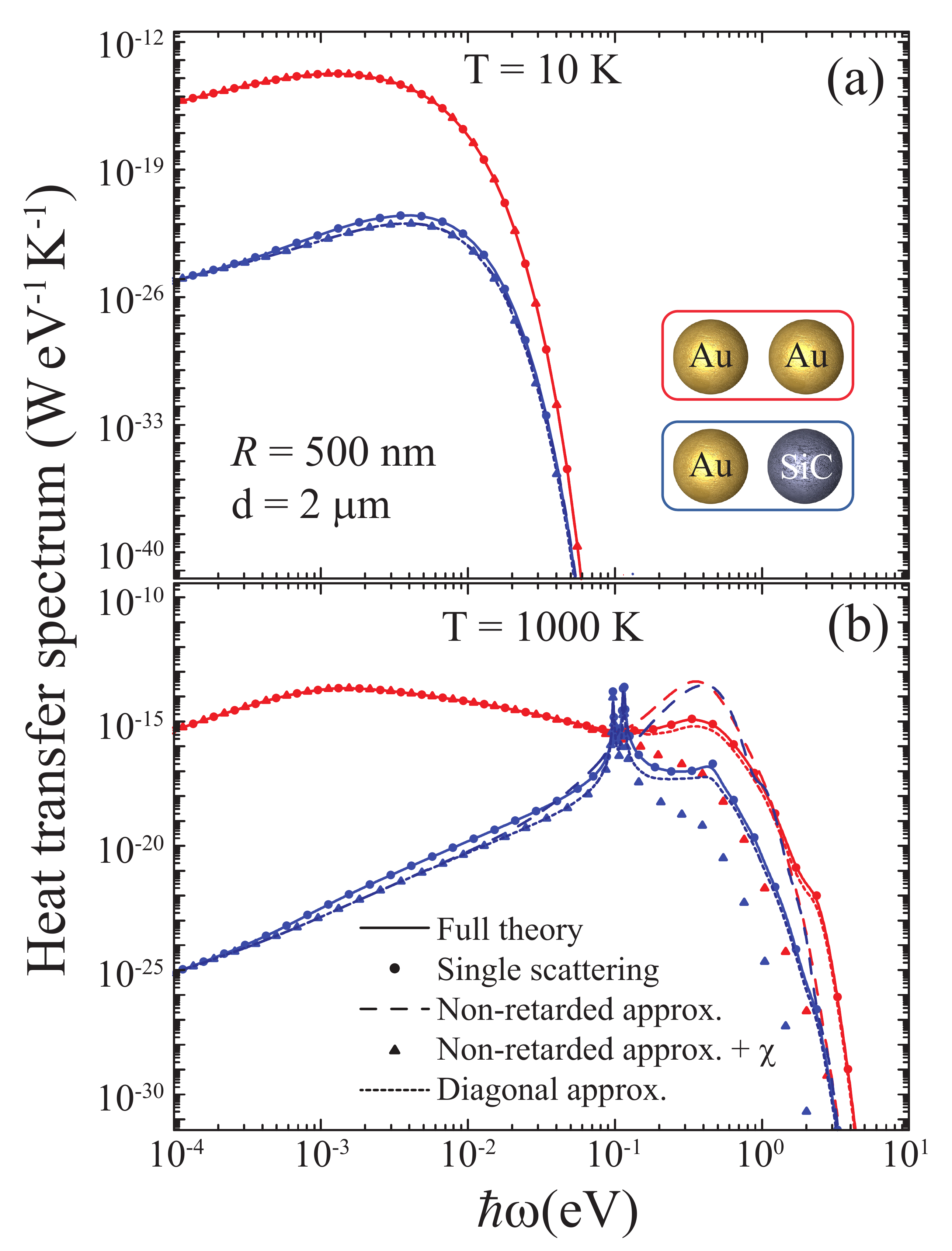}
\caption{(color online). Spectral dependence of the radiative HTC under the conditions of Fig.\ \ref{Fig3}(a) for two different temperatures, as shown by labels.} \label{Fig4}
\end{center}
\end{figure}

We study in Figs.\ \ref{Fig2}, \ref{Fig3}(a), and \ref{Fig4} the heat transfer coefficient (HTC) between two particles in a dimer when particle 1 is at the same temperature as the environment ($T_1=T_0=T$) and particle 2 is at a slightly different temperature ($T_2=T+\delta T$). The HTC to particle 1 is defined per unit of temperature difference as $\mathcal{P}_1/\delta T$. Under these conditions, only the terms $\Gamma^\nu_{2,\perp}$ and $\Gamma^\nu_{2,\parallel}$ contribute to the transfer. The results obtained from the above formalism (solid curves) are compared to several approximations consisting of neglecting multiple scattering between the particles (circles, calculated for $u_\nu=t_\nu=s=1$, $w_\nu=0$), retardation effects everywhere (dashed curves, $k=0$), retardation effects except in the particle response $\chi_j$ defined by Eq.\ (\ref{chi}) (triangles), or EMCTs (dotted curves).

The distance dependence of the HTC is analyzed in Fig.\ \ref{Fig2} for a homogeneous gold dimer and for a dimer formed by gold and SiC particles. As a first observation, we note that multiple scattering events can be safely neglected in all cases. In contrast, retardation causes a dramatic boost in the HTC, which increases with particle distance in both types of dimers. Additionally, the $|\alpha|^2$ term of Eq.\ (\ref{chi}) contributes with a uniform decreasing factor in SiC particles (see Appendix, Fig.\ 5). (Notice that the absorption cross section is proportional to $\chi_j$, whereas $\I\{\alpha_j\}$ describes absorption plus scattering, so the $|\alpha|^2$ term in $\chi_j$ is removing scattering strength that is not associated with absorption.) Finally, EMCTs terms introduce additional channels of inter-particle interaction, thus resulting in higher transfer rates compared to the diagonal approximation (consisting of only including electric-electric and magnetic-magnetic terms), particularly in the heterogeneous dimer [Fig.\ \ref{Fig1}(a)]. Figure\ \ref{Fig2}(c) clearly shows that the magnetic-magnetic terms are dominant in the homogeneous gold dimer, in agreement with previous predictions \cite{CLV08_1}, because metallic particles mainly contribute through magnetic polarization. This is unlike the heterogeneous cluster, in which magnetic-electric terms are dominant [Figure\ \ref{Fig2}(d)], thus picking up a dominant electric polarization from the SiC particle.  Incidentally, the HTCs from gold to SiC and from SiC to gold are nearly identical (see Appendix, Fig.\ 8).

Similar conclusions are extracted from the temperature dependence of the HTC, represented in Fig.\ \ref{Fig3}(a) for a small distance $d=2\,\mu$m$\ll\lambda_T=14\,\mu$m$-14$mm. Notice however the dramatic reduction in the transfer rate produced by retardation at high temperatures.

The full dependence on the particle temperatures for a vacuum at $T_0=0$ is studied in Fig.\ \ref{Fig3}(c,d). Interestingly, particle 1 gets cooled down (blue regions) even if particle 2 is at a higher temperature. This is due to radiation losses into the vacuum. However, particle 2 in the homogeneous cluster is rather efficient in transferring energy to particle 1 and compensating for radiation losses, so that the curve separating gains (red) from losses (blue) is closer to the $T_1=T_2$ line (dashed) in that dimer [Fig.\ \ref{Fig3}(c)].

It is useful to analyze the spectral contribution of different photon energies to the HTC. At a low temperature $T=10\,$K [Fig.\ \ref{Fig4}(a)], the exchange is dominated by low photon energies, for which the particles polarization show a featureless behavior and $\chi_j$ almost coincides with ${\rm Im}\{\alpha_j\}$ for the size of the particles under discussion (see  Appendix, Fig.\ 5). At high temperature $T=1000\,$K  [Fig.\ \ref{Fig4}(b)], optical phonons emerge as a sharp infrared (IR) feature in SiC and plasmons show up as a broader near-IR feature in gold particles. Retardation effects also increase with $T$, as the particles appear to be large in front of $\lambda_T$.

An important ingredient that is often overlooked in the analysis of heat transfer relates to how much energy is emitted into the surrounding vacuum. We analyze this in Figs.\ \ref{Fig2}(b) and \ref{Fig3}(b) by calculating the power escaping from a hotter particle 2. The calculation is done by reversing the particle labels, so that only the terms $\Gamma^\nu_{1,\perp}$ and $\Gamma^\nu_{1,\parallel}$ contribute in this case [see Eq.\ (\ref{Pall})]. For the temperature of Fig.\ \ref{Fig2}, just a small amount of the energy emanating from particle 2 ends up in particle 1. However, this fraction increases at lower temperatures [Figs.\ \ref{Fig3}(b)], until nearly complete heat transfer takes place below $\sim10\,$K in the homogeneous gold dimer. The fraction of heat transfer between the particles is thus very sensitive to temperature and particle distance (see  Appendix, Fig.\ 9).

\section{Concluding remarks}

Heat dissipation in nanostructured devices is becoming a limiting factor in the design of microchips and is expected to play a major role in nanoelectronics, nanophotonics, and photovoltaics. Radiative losses provide a convenient way of handling the excess of heat produced in these devices \cite{FZ06}. In this context, crossed magnetic-electric terms and radiative corrections as those described here produce modifications in the transfer rate by up to several orders of magnitude, which cannot be overlooked. An analysis of how much heat is released from a dimer into a cooler vacuum reveals a large dependence on composition and temperature (it is strongly suppressed at low temperatures and dominant in hot environments). Our results for gold-SiC dimers suggest the experimental exploration of these effects via, for example, in-vacuum particle levitation, or by attaching one of the particles to a nanoscale tip and the other one to an insulating substrate. As an interesting direction, we note that RHT can be strongly modified by the presence of additional mirrors and dielectrics that distort the exchanged electromagnetic fields. We further suggest the possibility of molding RHT down to the quantum regime by placing the particles in a resonant cavity. This directly connects to the proposed quantization of RHT \cite{BRG10}, similar to that observed in the conventional thermal conductance of narrow bridges \cite{SHW00}.

\begin{acknowledgments}
This work has been supported by the Spanish MICINN (MAT2010-14885 and Consolider NanoLight.es) and the European Commission (FP7-ICT-2009-4-248909-LIMA and FP7-ICT-2009-4-248855-N4E). A.M. acknowledges financial support through FPU from ME.
\end{acknowledgments}

\appendix

\begin{figure}
\begin{center}
\includegraphics[width=125mm,angle=0]{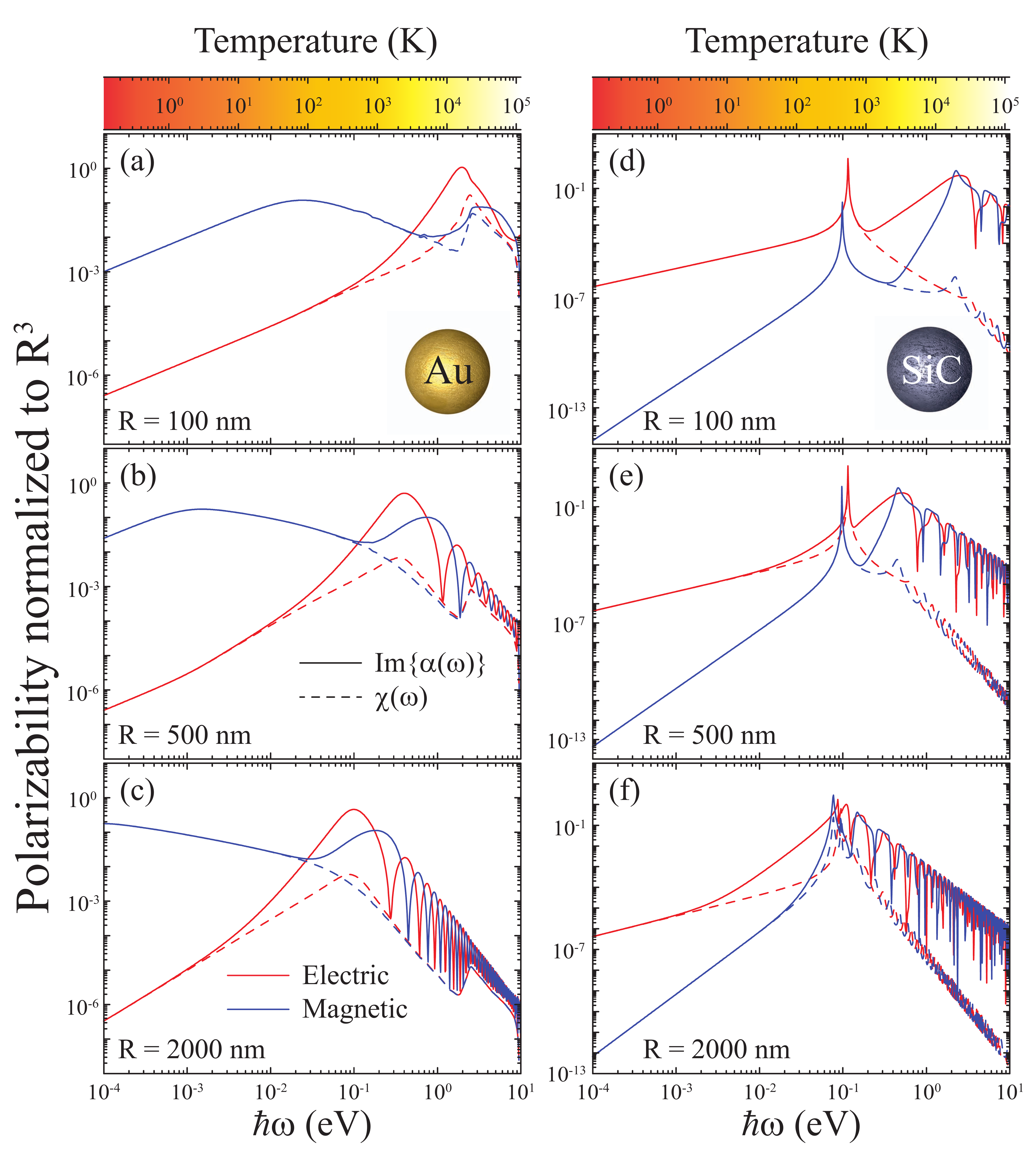}
\caption{Imaginary part of the polarizability ${\rm Im}\{\alpha^\nu\}$ (solid curves) and absorption coefficient $\chi^\nu={\rm Im}\{\alpha^\nu\}-(2k^3/3)|\alpha^\nu|^2$ (dashed curves) for gold (a)-(c) and SiC (d)-(f) spherical particles of different radius $R$, as a function of photon energy. Red and blue curves represent electric ($\nu=E$) and magnetic ($\nu=M$) components. The upper color scale shows the equivalent temperature $\hbar\omega/k_B$.}\label{Fig5}
\end{center}
\end{figure}

\section{Particle polarizability}

We obtain the polarizability of the homogeneous spherical particles under consideration from their dipolar Mie scattering coefficients as $\alpha_j^\nu=(3/2k^3)t_{j,1}^\nu$ ($\nu=E,M$), where $k=\omega/c$. This procedure automatically incorporates a number of retardation corrections in the polarizability. The Mie coefficients are given by the analytical expressions \cite{paper024}
\begin{align}
      t_l^M &= \frac{- j_l(\rho_0) \rho_1 j_l^\prime(\rho_1)
                    + \rho_0 j_l^\prime(\rho_0) j_l(\rho_1)}
                   {h_l^{(+)}(\rho_0) \rho_1 j_l^\prime(\rho_1)
                    -\rho_0  [h_l^{(+)}(\rho_0)]^\prime j_l(\rho_1)},
\nonumber \\
      t_l^E &= \frac{- j_l(\rho_0) [\rho_1 j_l(\rho_1)]^\prime
                    + \epsilon [\rho_0 j_l(\rho_0)]^\prime j_l(\rho_1)}
                   {h_l^{(+)}(\rho_0) [\rho_1 j_l(\rho_1)]^\prime
                    - \epsilon [\rho_0  h_l^{(+)}(\rho_0)]^\prime j_l(\rho_1)},
\nonumber
\end{align}
where $\rho_0=kR$, $\rho_1=kR\sqrt{\epsilon}$ with
${\rm Im}\{\rho_1\}>0$, $R$ is the particle radius, $\epsilon$ is its dielectric function, $j_l$ and $h_l^{(+)}$ are spherical Bessel and Hankel functions, and the prime denotes differentiation with respect to $\rho_0$
and $\rho_1$.

Figure \ref{Fig5} shows the imaginary part of the particle polarizability ${\rm Im}\{\alpha\}$ compared to the absorption factor $\chi={\rm Im}\{\alpha\}-2k^3/3|\alpha|^2$ for gold and SiC spheres of radius $R$ similar to those presented in Sec.\ III. As expected, $\chi$ approaches ${\rm Im}\{\alpha\}$ in the $kR\ll1$ limit (i.e., when retardation is negligible). However, ${\rm Im}\{\alpha\}$ and $\chi$ behave increasingly different as the energy goes up. The difference between these two functions increases with particle size, or equivalently, the threshold for retardation effects $\omega=c/R$ (e.g., 0.4\,eV for $R=500\,$nm) is lowered. Interestingly, the magnetic (electric) component is dominant in gold (SiC) particles at energies below $\sim0.1\,$eV, while plasmon (phonon) resonances take over for larger $\omega$.

\begin{figure}
\begin{center}
\includegraphics[width=85mm,angle=0]{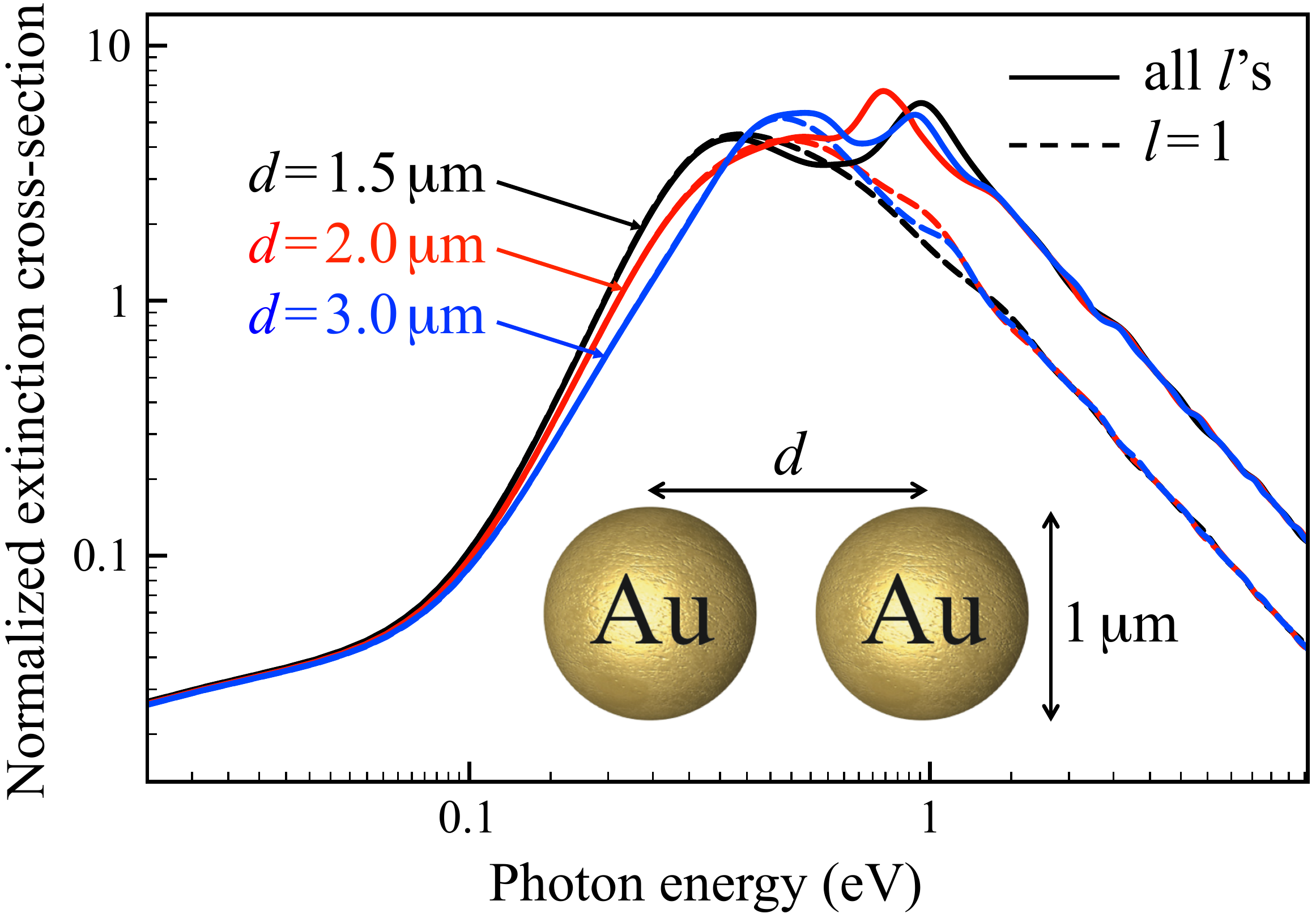}
\caption{Extinction cross-section spectra of gold dimers calculated in the dipolar approximation ($l=1$, dashed curves) and with full inclusion of all multipoles (solid curves) for different distances between particle centers $d$, as indicated by labels. The particles diameter is $1\,\mu$m. The cross section is normalized to the projected area of one particle.}\label{Fig6}
\end{center}
\end{figure}

\section{Validity of the dipolar approximation}

The dipolar approximation should be accurate in the limit of small particles compared to the light wavelength. The validity of this approximation is tested in Fig.\ \ref{Fig6} by comparing the extinction cross section of gold dimers calculated by representing the particles as dipoles (broken curves) or by including all multipoles (solid curves). The calculations are performed using a multiple elastic scattering of multipolar expansions (MESME) method \cite{paper024}. We obtain similar values of the cross section in both calculations for different particle separations down to $d=1.5\,\mu$m when the photon energy is smaller than $\hbar c/R\approx\,0.4$eV, where $R=500\,$nm is the particle radius. The complex plasmon features showing up at larger energies in the full calculation are not captured within the dipolar approximation. Our formalism is thus appropriate to describe dimers within this range of sizes and separations for temperatures below $\sim \hbar c/k_B R\approx4580\,$K.

\begin{figure}
\begin{center}
\includegraphics[width=100mm,angle=0]{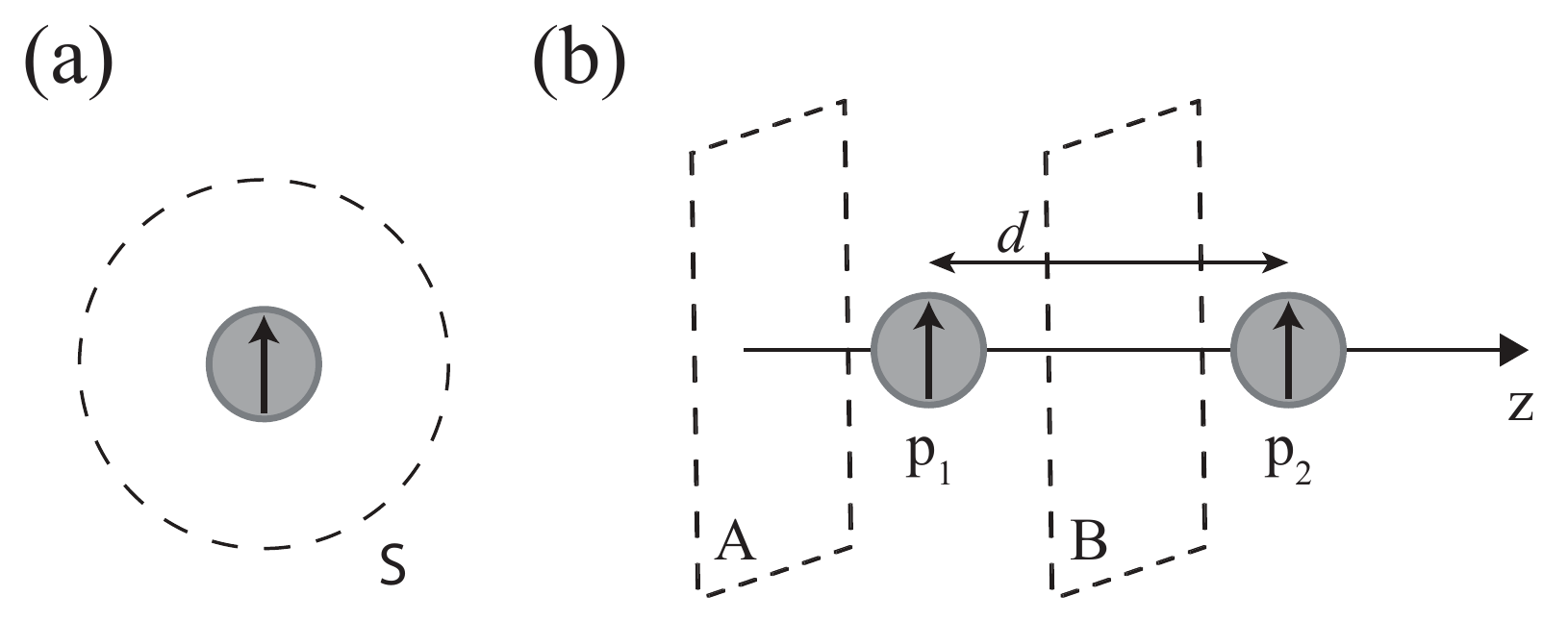}
\caption{Description of the geometry used in the derivation of Eqs.\ (1) and (2).}\label{Fig7}
\end{center}
\end{figure}

\section{Derivation of Eq.\ (1)}

We consider a spherical particle placed at the origin and of radius much smaller than the thermal wavelength $\lambda_T=2\pi\hbar c/k_BT$ [see Fig.~\ref{Fig7}(a)], so that it can be described by its electric and magnetic polarizabilities, $\alpha^E$ and $\alpha^M$, respectively. Under illumination by an external electromagnetic plane wave of frequency $\w$ and wave vector $\kk$, the total field (external plus induced) can be written
\begin{subequations}\label{w.1}
\begin{align}
\EEs(\rr,t)=&\EEs_0e^{i\kk\cdot\rr}e^{-i\w t} + \mathcal{G}(\rr,\w)\alpha^{\rm E}(\w)\EEs_0 e^{-i\w t} - \frac{1}{ik} \nabla \times \mathcal{G}(\rr,\w)\alpha^{\rm M}(\w)\HHs_0 e^{-i\w t} + {\rm c.c.}, \\
\HHs(\rr,t)=&\HHs_0e^{i\kk\cdot\rr}e^{-i\w t} + \mathcal{G}(\rr,\w)\alpha^{\rm M}(\w)\HHs_0 e^{-i\w t} + \frac{1}{ik} \nabla \times \mathcal{G}(\rr,\w)\alpha^{\rm E}(\w)\EEs_0 e^{-i\w t} + {\rm c.c.},
\end{align}\label{w.1}
\end{subequations}
where
\begin{align}
\mathcal{G}(\rr,\omega)=\left[k^2+\nabla\otimes\nabla\right]\frac{e^{ikr}}{r}
=\frac{e^{ikr}}{r^3}\left[(k^2r^2+ikr-1)-(k^2r^2+3ikr-3)\hat\rr\otimes\hat\rr\right], \label{Green.1}
\end{align}
is the electromagnetic Green tensor. The work exerted on the particle by the external electromagnetic field can be obtained from the Poynting vector flux across a spherical surface $S$ of radius $r$ centered at the origin as
\begin{align}\label{w.2}
\mathcal{P}^{\rm field}_{1}=\frac{c}{4\pi}\,r^2 \int_S d\Omega \left[\EEs(\rr,t)\times \HHs(\rr,t)\right]\cdot \hat{\rr}.
\end{align}
Now, inserting Eqs.~(\ref{w.1}) into Eq.~(\ref{w.2}), we readily obtain the expression
\begin{align}
\mathcal{P}^{\rm field}_{1}=\w \left[-i\alpha^{\rm E}-\frac{2k^3}{3}\left|\alpha^{\rm E}\right|^2\right]\left|\EEs_0\right|^2+\w \left[-i\alpha^{\rm M}-\frac{2k^3}{3}\left|\alpha^{\rm M}\right|^2\right]\left|\HHs_0\right|^2+ {\rm c.c.}, \label{P1.1}
\end{align}
where we have only retained stationary terms. Equation\ (\ref{P1.1}) is the monochromatic version of Eq.\ (1).

\section{Derivation of Eq.\ (2)}

We now consider two neighboring spherical particles placed at the origin and at a distance $d$ from the origin along the $z$ axis, respectively, as shown in Fig.~\ref{Fig7}(b). We assume small particles compared to both $\lambda_T$ and $d$, so that we can describe them as point electric and magnetic dipoles, $\pp_j$ and $\mm_j$, respectively, where $j=1,2$ labels the particles. We calculate the power absorbed by particle 1 from the Poynting vector flux in the direction towards the particle across two infinite planes A and B normal to $\hat{\bf z}$, as shown in Fig.~\ref{Fig7}(b). More precisely,
\begin{align}
\mathcal{P}^{\rm dip}_{1}=\frac{c}{4\pi} \left(\int_{A}-\int_B\right) d\RR \left[\EEs(\rr,t)\times \HHs(\rr,t)\right]\cdot \hat{\zz}. \label{x.1}
\end{align}
Taking advantage of this geometry, the electromagnetic field created by the electric and magnetic particle dipoles can be written
\begin{subequations}
\begin{align}
\EEs(\rr,t)=&\frac{ie^{-i\w t}}{2\pi}\sum_{i=1,2}\int \frac{d\QQ}{q} \left[\left(k^2+\nabla\otimes\nabla\right)\pp_i - ik \nabla\times\mm_i \right]e^{i\QQ\cdot\RR}e^{iq|z-z_i|} + {\rm c.c.}, \\
\HHs(\rr,t)=&\frac{ie^{-i\w t}}{2\pi}\sum_{i=1,2}\int \frac{d\QQ}{q} \left[\left(k^2+\nabla\otimes\nabla\right)\mm_i + ik \nabla\times\pp_i \right]e^{i\QQ\cdot\RR}e^{iq|z-z_i|} + {\rm c.c.},
\end{align}\label{x.2}
\end{subequations}
where we have used the relation
\begin{align}
\frac{e^{ikr}}{r}=\frac{i}{2\pi}\int \frac{d\QQ}{q} e^{i\QQ\cdot\RR}e^{iq|z-z_i|}, \nonumber
\end{align}
the integrals are extended over wave vectors along $x$ and $y$ directions, and $q=\sqrt{k^2-Q^2}$ ($q=i\sqrt{Q^2-k^2}$) is the wave vector along $z$ for $k\ge Q$ ($k<Q$). Finally, inserting Eqs.~(\ref{x.2}) into Eq.~(\ref{x.1}) we obtain, after some straightforward algebra,
\begin{align}
\mathcal{P}^{\rm dip}_{1}={}&i\w\A \left(p_{1x}^*p_{2x}+p_{1y}^*p_{2y}+m_{1x}^*m_{2x}+m_{1y}^*m_{2y}\right)+i\w\B\left(p_{1z}^*p_{2z}+m_{1z}^*m_{2z}\right)\nonumber \\ &+i\w\C\left(-p_{1x}^*m_{2y}+p_{1y}^*m_{2x}+m_{1x}^*p_{2y}-m_{1y}^*p_{2x}\right)-\frac{2\w^4}{3c^3}\left(\left|\pp_1\right|^2+\left|\mm_1\right|^2\right) + {\rm c.c.}, \nonumber
\end{align}
which is the monochromatic version of Eq.\ (2).

\section{Fluctuation-dissipation theorem}

The fluctuation-dissipation theorem (FDT) was first formulated by Nyquist \cite{N1928} and later proved by Callen and Welton \cite{CW1951}. It relates the fluctuations of the product of two operators to the dissipation expressed as the imaginary part of their response function. Next, we reproduce for convenience a simple derivation given elsewhere \cite{paper157} with a notation suited to the analysis presented in this paper.

Consider a perturbation Hamiltonian
\begin{eqnarray}
\Hh'(t)=-\varphi(t)\qh(t),\nonumber
\end{eqnarray}
where $\varphi(t)$ is a time-dependent function, $\qh(t)$ is an operator in the Heisenberg picture, related to its Schr\"odinger representation $\qh_S$ through $\qh(t)=\exp(i\Hh_0t/\hbar)\qh_S\exp(-i\Hh_0t/\hbar)$, and $\Hh_0$ is the unperturbed Hamiltonian. In the Heisenberg representation, Schr\"odinger's equation becomes $\Hh'|\phi\rangle=i\hbar\partial|\phi\rangle/\partial t$, and we have $|\phi\rangle=\exp(i\Hh_0t/\hbar)|\phi_S\rangle$.

Under the condition $\Hh'(-\infty)=0$, the eigenstates of the total Hamiltonian become
\begin{eqnarray}
|\phi_m(t)\rangle&=&|m\rangle-\frac{i}{\hbar}\int_{-\infty}^tdt'H'(t')|\phi_m(t')\rangle\nonumber\\
&\approx&|m\rangle-\frac{i}{\hbar}\int_{-\infty}^tdt'H'(t')|m\rangle,\nonumber
\end{eqnarray}
where the last line corresponds to first-order perturbation theory, and $|m\rangle$ is a state of the unperturbed Hamiltonian with energy $E_m$ (i.e., $\Hh_0|m\rangle=E_m|m\rangle$).

The expected value of another operator $\ph(t)$ is simply given by
\begin{eqnarray}
\left\langle\ph(t)\right\rangle&=&\frac{1}{Z}\sum_me^{-E_m/k_BT}\left\langle\phi_m(t)|\ph(t)|\phi_m(t)\right\rangle\nonumber\\
&\approx&\frac{1}{Z}\sum_me^{-E_m/k_BT}\left[\left\langle m|\ph(t)|m\right\rangle+\frac{i}{\hbar}\int_{-\infty}^tdt'\varphi(t')\left\langle m|[\ph(t),\qh(t')]|m\right\rangle\right], \label{eq1}
\end{eqnarray}
where
\begin{eqnarray}
Z=\sum_me^{-E_m/k_BT}\nonumber
\end{eqnarray}
is the partition function at temperature $T$. The first term in Eq.\ (\ref{eq1}) reduces to $\langle m|\ph(t)|m\rangle=\langle m|\ph_S|m\rangle$, and from here, we can recast (\ref{eq1}) as
\begin{eqnarray}
\left\langle\delta\ph(t)\right\rangle\equiv\left\langle\ph(t)-\ph(-\infty)\right\rangle=\int dt'\chi(t-t')\varphi(t'), \nonumber
\end{eqnarray}
where
\begin{eqnarray}
\chi(t-t')=\frac{i}{\hbar}\theta(t-t')\frac{1}{Z}\sum_me^{-E_m/k_BT}\left\langle m|[\ph(t),\qh(t')]|m\right\rangle \label{eq2}
\end{eqnarray}
is a susceptibility. Now, using the closure relation $|n\rangle\langle n|=I$, we can write
\begin{eqnarray}
\left\langle m|[\ph(t),\qh(t')]|m\right\rangle=\sum_n\left[\left\langle m|\ph_S|n\right\rangle\left\langle n|\qh_S|m\right\rangle e^{i(E_m-E_n)(t-t')/\hbar}-\left\langle m|\qh_S|n\right\rangle\left\langle n|\ph_S|m\right\rangle e^{-i(E_m-E_n)(t-t')/\hbar}\right]. \nonumber
\end{eqnarray}
Inserting this back into Eq.\ (\ref{eq2}), taking the time Fourier transform to work in frequency space, and using the identity
\begin{eqnarray}
\int_0^\infty dt\;e^{i\Delta t}=\frac{i}{\Delta+i0^+}, \nonumber
\end{eqnarray}
we find
\begin{eqnarray}
\chi(\omega)&=&\int dt\;\chi(t)\;e^{i\omega t} \nonumber\\
&=&\frac{-1}{Z}\sum_{m,n}\left\langle m|\ph_S|n\right\rangle\left\langle n|\qh_S|m\right\rangle\frac{e^{-E_m/k_BT}-e^{-E_n/k_BT}}{\hbar\omega+E_m-E_n+i0^+}. \nonumber
\end{eqnarray}
Incidentally, the zero-temperature susceptibility reads
\begin{eqnarray}
\chi(\omega)\;\;_{\overrightarrow{T\rightarrow0}}\;\;\;-\sum_m\left[\frac{\left\langle 0|\ph_S|m\right\rangle\left\langle m|\qh_S|0\right\rangle}{\hbar\omega+E_0-E_m+i0^+}
-\frac{\left\langle 0|\qh_S|m\right\rangle\left\langle m|\ph_S|0\right\rangle}{\hbar\omega+E_m-E_0+i0^+}\right]. \nonumber
\end{eqnarray}
Finally, the dissipation associated with $\chi$ can be written
\begin{eqnarray}
{\rm Im}\left\{\chi(\omega)\right\}=\left(1-e^{-\hbar\omega/k_BT}\right)\frac{\pi}{Z}\sum_{m,n}e^{-E_m/k_BT}\left\langle m|\ph_S|n\right\rangle\left\langle n|\qh_S|m\right\rangle\;\delta(\hbar\omega+E_m-E_n). \label{eqdis}
\end{eqnarray}

Similarly, we can write the average over fluctuations as
\begin{eqnarray}
S(t-t')\equiv\left\langle\ph(t)\qh(t')\right\rangle=\frac{1}{Z}\sum_{m,n}e^{-E_m/k_BT}\;e^{i(E_m-E_n)(t-t')/\hbar}\left\langle m|\ph_S|n\right\rangle\left\langle n|\qh_S|m\right\rangle
=\int\frac{d\omega}{2\pi} S(\omega) e^{-i\omega(t-t')}, \nonumber
\end{eqnarray}
where
\begin{eqnarray}
S(\omega)=\frac{2\pi\hbar}{Z}\sum_{m,n}e^{-E_m/k_BT}\;\left\langle m|\ph_S|n\right\rangle\left\langle n|\qh_S|m\right\rangle\;\delta(\hbar\omega+E_m-E_n). \label{eqflu}
\end{eqnarray}

The relation between $S(\omega)$ and ${\rm Im}\left\{\chi(\omega)\right\}$ that one obtains by comparing Eqs.\ (\ref{eqdis}) and (\ref{eqflu}) constitutes the general form of the fluctuation-dissipation theorem:
\begin{eqnarray}
S(\omega)=2\hbar\;[n(\omega)+1]\;{\rm Im}\left\{\chi(\omega)\right\}, \label{Sw}
\end{eqnarray}
where
\begin{eqnarray}\label{BEd}
n(\omega)=\frac{1}{e^{\hbar\omega/k_BT}-1},
\end{eqnarray}
is the Bose-Einstein distribution function.

We can formulate a more useful relation by noticing that $\left\langle\ph(t)\qh(t')\right\rangle$ is a function of $t-t'$, so that its double Fourier transform satisfies
\begin{eqnarray}
\left\langle\ph(\omega)\qh(\omega')\right\rangle=\int\;dt\,dt'\;e^{i\omega t+i\omega't'} S(t-t')=\int\;d\tau\;e^{i\omega\tau}\;S(\tau)\;\int dt'\;e^{i(\omega+\omega')t'}=2\pi\;\delta(\omega+\omega')\;S(\omega), \nonumber
\end{eqnarray}
and from here we find the expression
\begin{eqnarray}
\left\langle\ph(\omega)\qh(\omega')\right\rangle&=&4\pi\hbar\;[n(\omega)+1]\;{\rm Im}\left\{\chi(\omega)\right\}\;\delta(\omega+\omega'). \label{fdt1}
\end{eqnarray}
Proceeding as above, the Fourier transform of the fluctuation $\left\langle\qh(t')\ph(t)\right\rangle$ reads $\left\langle\qh(\omega')\ph(\omega)\right\rangle=\exp(-\hbar\omega/k_BT)\,S(\omega)$, which together with (\ref{Sw}) leads to
\begin{eqnarray}
\left\langle\qh(\omega')\ph(\omega)\right\rangle&=&4\pi\hbar\;n(\omega)\;{\rm Im}\left\{\chi(\omega)\right\}\;\delta(\omega+\omega'). \label{fdt2}
\end{eqnarray}
Finally, it should be noticed that $\ph(\omega)\qh(\omega')$ is not an observable in general, but the symmetrized product is Hermitian, and therefore, an observable. From Eqs.\ (\ref{fdt1}) and (\ref{fdt2}), we find
\begin{eqnarray}
\frac{1}{2}\left\langle\ph(\omega)\qh(\omega')+\qh(\omega')\ph(\omega)\right\rangle=4\pi\hbar\left[n(\omega)+\frac{1}{2}\right]{\rm Im}\left\{\chi(\omega)\right\}\delta(\omega+\omega').\label{fdt3}
\end{eqnarray}
Equations\ (\ref{fdt1})-(\ref{fdt3}) are general forms of the FDT.

\subsection{FDT for induced dipole fluctuations}

We apply the above general expressions of the FDT to dipole-dipole fluctuations, with the identifications
\begin{eqnarray}
\ph(t)&\rightarrow&p_i(t), \nonumber\\
\qh(t)&\rightarrow&p_j(t), \nonumber\\
\chi(t)&\rightarrow&\alpha_{ij}(t), \nonumber\\
\varphi(t)&\rightarrow&E_j(t), \nonumber
\end{eqnarray}
where $p_i$ and $p_j$ are components of the dipole moment along directions $i$ and $j$, respectively, $E_j$ is the electric field along $j$ at the position of the dipole, and $\alpha_{ij}$ is the $(i,j)$ component of the polarizability tensor. The interaction Hamiltonian is $\Hh'(t)=-E_j(t)p_j(t)$, where $E_j$ is regarded as a time-dependent function and $q_j$ as an operator. The susceptibility acts in frequency space according to $\left\langle\delta p_i(\omega)\right\rangle=\alpha_{ij}(\omega)E_j(\omega)$. With these substitutions, the FDT [Eqs.\ (\ref{fdt1})-(\ref{fdt3})] takes the forms
\begin{eqnarray}
\left\langle p_i(\omega)p_j(\omega')\right\rangle&=&4\pi\hbar\;\delta(\omega+\omega')\;{\rm Im}\left\{\alpha_{ij}(\omega)\right\}\;[n(\omega)+1], \nonumber\\
\left\langle p_j(\omega')p_i(\omega)\right\rangle&=&4\pi\hbar\;\delta(\omega+\omega')\;{\rm Im}\left\{\alpha_{ij}(\omega)\right\}\;n(\omega), \nonumber\\
\frac{1}{2}\left\langle p_i(\omega)p_j(\omega')+p_j(\omega')p_i(\omega)\right\rangle&=&4\pi\hbar\;\delta(\omega+\omega')\;{\rm Im}\left\{\alpha_{ij}(\omega)\right\}\;[n(\omega)+\frac{1}{2}]. \label{FDTp}
\end{eqnarray}
From this formulation, the dissipation is found to be proportional to ${\rm Im}\{\alpha\}$. However, this leads to unphysical results in non-absorbing particles. For example, in spheres, non-vanishing values of ${\rm Im}\{\alpha\}=(2k^3/3)|\alpha|^2$ arise from retardation corrections. Direct application of Eq.\ (\ref{FDTp}) to that case predicts an unphysical amount of absorption. In order to correct for this, we use a modified version of the FDT for spherical particles ($\alpha_{ij}=\delta_{ij}\alpha$),
\begin{equation}
\frac{1}{2}\left\langle p_i(\omega)p_j(\omega')+p_j(\omega')p_i(\omega)\right\rangle=4\pi\hbar\;\delta(\omega+\omega')\;\chi(\omega)\;\delta_{ij}\;[n(\omega)+\frac{1}{2}], \label{FDTp}
\end{equation}
where $\chi={\rm Im}\{\alpha\}-(2k^3/3)|\alpha|^2$. The above derivation can be straightforwardly applied to magnetic dipole fluctuations.

\section{Fluctuation-dissipation theorem for electromagnetic fields}

Although it is possible to obtain an expression similar to Eq.\  (\ref{FDTp}) for the FDT corresponding to the fluctuations of the electromagnetic field following the above formalism \cite{paper157}, we present next an alternative derivation based upon the direct evaluation of electromagnetic quantum operator correlations.

\subsection{Electric field fluctuations}

The electric field quantum operator is defined as \cite{L1983}
\begin{align}\nonumber
\EE(\rr,\omega)=\EE^{(+)}(\rr,\omega)+\EE^{(-)}(\rr,\omega),
\end{align}
where
\begin{subequations}
\begin{align}
\EE^{(+)}(\rr,\omega)&=i\sqrt{\frac{2\pi\hbar\omega}{V}}\ee_{\kk,\sigma}\ag_{\kk,\sigma}e^{i\kk\cdot\rr}, \\
\EE^{(-)}(\rr,\omega)&=-i\sqrt{\frac{2\pi\hbar\omega}{V}}\ee_{\kk,\sigma}\ag^{+}_{\kk,\sigma}e^{-i\kk\cdot\rr}.
\end{align} \label{e.2}
\end{subequations}
Here, $\kk$ represents the wave vector, $\sigma$ is the polarization state, $\ee_{\kk',\sigma'}$ is the polarization vector, and $\ag^{+}_{\kk,\sigma}$ ($\ag^{+}_{\kk,\sigma}$) is the photon annihilation (creation) operator, which acts over photon states as
\begin{align}
\ag^{+}_{\kk',\sigma'}\ket{n_{\kk,\sigma}}&=(n_{\kk,\sigma}+1)\delta_{\kk,\kk'}\delta_{\sigma,\sigma'}\ket{(n+1)_{\kk,\sigma}}, \nonumber \\
\ag_{\kk',\sigma'}\ket{n_{\kk,\sigma}}&=n_{\kk,\sigma}\delta_{\kk,\kk'}\delta_{\sigma,\sigma'}\ket{(n-1)_{\kk,\sigma}}. \nonumber
\end{align}
Using these expressions, we can calculate the correlation function of the electric field for normal ordering,
\begin{align}
\left\langle\EE^{(-)}(\rr_1,\omega')\otimes\EE^{(+)}(\rr_2,\omega'')\right\rangle&=\sum_{\kk,\sigma,n}\bra{n_{\kk,\sigma}}\EE^{(-)}(\rr_1,\omega')\otimes\EE^{(+)}(\rr_2,\omega'')\ket{n_{\kk,\sigma}}
\label{e.4} \\ &=\sum_{n}\frac{V}{(2\pi)^3c^3}\int d\Omega \sum_{\sigma}\int d\omega \omega^2 \bra{n_{\kk,\sigma}}\EE^{(-)}(\rr_1,\omega')\otimes\EE^{(+)}(\rr_2,\omega'')\ket{n_{\kk,\sigma}}, \nonumber
\end{align}
where we assume a continuum of states and $d\Omega$ denotes an element of solid angle in $\kk$ space. The matrix element in Eq.\ (\ref{e.4}) reduces to
\begin{align}\nonumber
&\bra{n_{\kk,\sigma}}\EE^{(-)}(\rr_1,\omega')\otimes\EE^{(+)}(\rr_2,\omega'')\ket{n_{\kk,\sigma}}=\frac{2\pi\hbar\omega}{V}\ee_{\kk,\sigma}\otimes\ee_{\kk,\sigma}e^{-i\kk\cdot\rr}n_{\kk,\sigma}2\pi\delta(\omega-\omega')2\pi\delta(\omega-\omega''),
\end{align}
where $\rr=\rr_1-\rr_2$. Now, taking the average over all possible wave vectors and polarizations, we obtain
\begin{align}
&\frac{V}{(2\pi)^3c^3}\int d\Omega \sum_{\sigma}\int d\omega \omega^2\bra{n_{\kk,\sigma}}\EE^{(-)}(\rr_1,\omega')\otimes\EE^{(+)}(\rr_2,\omega'')\ket{n_{\kk,\sigma}}=\hbar\int d\Omega \sum_{\sigma}k^3 \ \ee_{\kk,\sigma}\otimes\ee_{\kk,\sigma}e^{-i\kk\cdot\rr}n_{\kk,\sigma}\delta(\omega'-\omega''). \nonumber
\end{align}
We use polarization vectors $\ee_{p}=\hat{\bf \theta}$ and $\ee_{s}=\hat{\bf \phi}$, which together with the wave vector $\kk$ form an orthogonal basis set, such that
\begin{align}
\sum_{\sigma}\ee_{\sigma}\otimes\ee_{\sigma}=\II-\hat\kk\otimes\hat\kk. \label{etimee}
\end{align}
Moreover, we have
\begin{align}\nonumber
\int d\Omega\,e^{\pm i\kk\cdot\rr} = 2\pi \int_{0}^{\pi}d\theta \sin(\theta)e^{\pm ikr\cos(\theta)} = 2\pi \frac{e^{ikr}-e^{-ikr}}{ikr}
\end{align}
and
\begin{align}\nonumber
\int d\Omega\,\kkg\otimes\kkg e^{\pm i\kk\cdot\rr}= -\frac{2\pi}{k^2} \nabla\otimes\nabla \left(\frac{e^{ikr}-e^{-ikr}}{ikr} \right).
\end{align}
Using all these results in Eq.~(\ref{e.4}) and assuming that the photons are at thermal equilibrium, we find
\begin{align}
\left\langle\EE^{(-)}(\rr_1,\omega')\otimes\EE^{(+)}(\rr_2,\omega'')\right\rangle=4\pi\hbar\delta(\omega'-\omega'')\,\mbox{Im}\left\{\mathcal{G}(\rr_1-\rr_2,\omega)\right\}\,n(\omega), \label{eeq1}
\end{align}
where $\mathcal{G}$ is the Green tensor defined in Eq.\ (\ref{Green.1}) and $n(\w)$ is the Bose-Einstein distribution function given by Eq.~(\ref{BEd}).

Following exactly the same procedure, we can calculate the correlation function of the electric field for antinormal ordering,
\begin{align}
\left\langle\EE^{(+)}(\rr_1,\omega')\otimes\EE^{(-)}(\rr_2,\omega'')\right\rangle&=\sum_{\kk,\sigma,n}\bra{n_{\kk,\sigma}}\EE^{(-)}(\rr_1,\omega')\otimes\EE^{(+)}(\rr_2,\omega'')\ket{n_{\kk,\sigma}}
\nonumber \\ &=\sum_{n}\frac{V}{(2\pi)^3c^3}\int d\Omega \sum_{\sigma}\int d\omega \omega^2 \bra{n_{\kk,\sigma}}\EE^{(+)}(\rr_1,\omega')\otimes\EE^{(-)}(\rr_2,\omega'')\ket{n_{\kk,\sigma}}, \nonumber
\end{align}
which reduces to an expression similar to Eq.\ (\ref{eeq1}), except that $n(\omega)$ has to be substituted by $n(\omega)+1$:
\begin{align}\nonumber
\left\langle\EE^{(+)}(\rr_1,\omega)\otimes\EE^{(-)}(\rr_2,\omega')\right\rangle=4\pi\hbar\delta(\omega-\omega')\, \mbox{Im}\left\{\mathcal{G}(\rr_1-\rr_2,\omega)\right\}\,[n(\omega)+1].
\end{align}

In a semiclassical framework such as the one used here (i.e., when the electromagnetic fields are classical magnitudes), the field fluctuations correspond to the symmetrized ordering, and therefore we have
\begin{align}\nonumber
\left\langle\EEs^{*}(\rr_1,\omega)\otimes\EEs(\rr_2,\omega')\right\rangle={}&\frac1{2}\left\langle\EE^{(-)}(\rr_1,\omega)\otimes\EE^{(+)}(\rr_2,\omega')+\EE^{(+)}(\rr_2,\omega)\otimes\EE^{(-)}(\rr_1,\omega')\right\rangle\\ \nonumber={}&4\pi\hbar\delta(\omega-\omega')\,\mbox{Im}\left\{\mathcal{G}(\rr_1-\rr_2,\omega)\right\}\,[n(\omega)+\frac{1}{2}].
\end{align}

\subsection{Magnetic field fluctuations}

The Maxwell-Faraday law \cite{J99},
\begin{align}\nonumber
-\frac1{c}\frac{\partial}{\partial t}\HH(\rr,t)=\nabla\times\EE(\rr,t),
\end{align}
allows us to obtain the magnetic field operator from the electric field [see Eqs.~(\ref{e.2})]:
\begin{align}\nonumber
\HH(\rr,\omega)=\HH^{(+)}(\rr,\omega)+\HH^{(-)}(\rr,\omega),
\end{align}
where
\begin{align}
\HH^{(+)}(\rr,\omega)&=i\sqrt{\frac{2\pi\hbar\omega}{V}}\left(\kkg\times\ee_{\kk,\sigma}\right)\ag_{\kk,\sigma}e^{i\kk\cdot\rr}, \nonumber \\
\HH^{(-)}(\rr,\omega)&=-i\sqrt{\frac{2\pi\hbar\omega}{V}}\left(\kkg\times\ee_{\kk,\sigma}\right)\ag^{+}_{\kk,\sigma}e^{-i\kk\cdot\rr}. \nonumber
\end{align}
In way similar to the electrical field fluctuations, the correlation function of the magnetic field is given by
\begin{align}
\left\langle\HH^{(-)}(\rr_1,\omega')\otimes\HH^{(+)}(\rr_2,\omega'')\right\rangle&=\sum_{\kk,\sigma,n}\bra{n_{\kk,\sigma}}\HH^{(-)}(\rr_1,\omega')\otimes\HH^{(+)}(\rr_2,\omega'')\ket{n_{\kk,\sigma}}
\nonumber \\ &=\sum_{n}\frac{V}{(2\pi)^3c^3}\int d\Omega \sum_{\sigma}\int d\omega \omega^2 \bra{n_{\kk,\sigma}}\HH^{(-)}(\rr_1,\omega')\otimes\HH^{(+)}(\rr_2,\omega'')\ket{n_{\kk,\sigma}} \nonumber
\end{align}
for normal ordering, and
\begin{align}
\left\langle\HH^{(+)}(\rr_1,\omega')\otimes\HH^{(-)}(\rr_2,\omega'')\right\rangle&=\sum_{\kk,\sigma,n}\bra{n_{\kk,\sigma}}\HH^{(+)}(\rr_1,\omega')\otimes\HH^{(-)}(\rr_2,\omega'')\ket{n_{\kk,\sigma}}
\nonumber \\ &=\sum_{n}\frac{V}{(2\pi)^3c^3}\int d\Omega \sum_{\sigma}\int d\omega \omega^2\bra{n_{\kk,\sigma}}\HH^{(+)}(\rr_1,\omega')\otimes\HH^{(-)}(\rr_2,\omega'')\ket{n_{\kk,\sigma}}, \nonumber
\end{align}
for antinormal ordering. From here, the calculation of the correlation functions is exactly the same as in the electric field. Incidentally, the sum over polarizations now becomes
\begin{align}\nonumber
\sum_{\sigma}\left(\hat\kkg\times\ee_{\kk,\sigma}\right)\otimes\left(\hat\kkg\times\ee_{\kk,\sigma}\right)=\sum_{\sigma}\ee_{\kk,\sigma}\otimes\ee_{\kk,\sigma},
\end{align}
so that we are again concerned with the same sum as in Eq.\ (\ref{etimee}), and therefore, the final result is the same as in the electric case:
\begin{align}\nonumber
\left\langle\HH^{(-)}(\rr_1,\omega)\otimes\HH^{(+)}(\rr_2,\omega')\right\rangle=4\pi\hbar\delta(\omega-\omega')\, \mbox{Im}\left\{\mathcal{G}(\rr_1-\rr_2,\omega)\right\}\,n(\omega),
\end{align}
\begin{align}\nonumber
\left\langle\HH^{(+)}(\rr_1,\omega)\otimes\HH^{(-)}(\rr_2,\omega')\right\rangle=4\pi\hbar\delta(\omega-\omega')\, \mbox{Im}\left\{\mathcal{G}(\rr_1-\rr_2,\omega)\right\}\,[n(\omega)+1],
\end{align}
\begin{align}\nonumber
\frac1{2}\left\langle\HH^{(-)}(\rr_1,\omega)\otimes\HH^{(+)}(\rr_2,\omega')+\HH^{(+)}(\rr_2,\omega)\otimes\HH^{(-)}(\rr_1,\omega')\right\rangle=4\pi\hbar\delta(\omega-\omega')\, \mbox{Im}\left\{\mathcal{G}(\rr_1-\rr_2,\omega)\right\}\,[n(\omega)+\frac1{2}].
\end{align}

\subsection{Electric-magnetic field fluctuations}

The correlation function of the electric-magnetic fields with normal ordering is
\begin{align}
\left\langle\EE^{(-)}(\rr_1,\omega')\otimes\HH^{(+)}(\rr_2,\omega'')\right\rangle&=\sum_{\kk,\sigma,n}\bra{n_{\kk,\sigma}}\EE^{(-)}(\rr_1,\omega')\otimes\HH^{(+)}(\rr_2,\omega'')\ket{n_{\kk,\sigma}}
\label{e.26} \\ &=\sum_{n}\frac{V}{(2\pi)^3c^3}\int d\Omega \sum_{\sigma}\int d\omega \omega^2 \bra{n_{\kk,\sigma}}\EE^{(-)}(\rr_1,\omega')\otimes\HH^{(+)}(\rr_2,\omega'')\ket{n_{\kk,\sigma}}. \nonumber
\end{align}
From the Maxwell-Faraday law, we find
\begin{subequations}
\begin{align}
\HH^{(+)}(\rr,\omega'')&=\frac1{ik}\nabla\times\EE^{(+)}(\rr,\omega''), \label{e.27a} \\
\HH^{(-)}(\rr,\omega'')&=-\frac1{ik}\nabla\times\EE^{(-)}(\rr,\omega'').
\label{e.27b}
\end{align} \label{e.27}
\end{subequations}
Using Eq.\ (\ref{e.27a}), we can rewrite Eq.~(\ref{e.26}) as
\begin{align}
\left\langle\EE^{(-)}(\rr_1,\omega')\otimes\HH^{(+)}(\rr_2,\omega'')\right\rangle=\frac{1}{ik}\nabla_2\times\left\{\sum_{n}\frac{V}{(2\pi)^3c^3}\int d\Omega \sum_{\sigma}\int d\omega \omega^2 \bra{n_{\kk,\sigma}}\EE^{(-)}(\rr_1,\omega')\otimes\EE^{(+)}(\rr_2,\omega'')\ket{n_{\kk,\sigma}}\right\}, \nonumber
\end{align}
and from here, we obtain
\begin{align}\label{e.28}
\left\langle\EE^{(-)}(\rr_1,\omega)\otimes\HH^{(+)}(\rr_2,\omega')\right\rangle=4\pi\hbar\delta(\omega-\omega')\,\frac{1}{ik}\mbox{Im}\left\{\nabla_2\times \mathcal{G}(\rr_1-\rr_2,\omega)\right\}\,n(\omega).
\end{align}
Likewise, for antinormal ordering, using Eq.~(\ref{e.27b}), we find
\begin{align}\label{e.29}
\left\langle\EE^{(+)}(\rr_1,\omega)\otimes\HH^{(-)}(\rr_2,\omega')\right\rangle=-4\pi\hbar\delta(\omega-\omega')\,\frac{1}{ik}\mbox{Im}\left\{\nabla_2\times \mathcal{G}(\rr_1-\rr_2,\omega)\right\}\,[n(\omega)+1].
\end{align}

\subsection{Magnetic-electric field fluctuations}

The correlation function of the magnetic-electric fields for normal ordering can be written
\begin{align}
\left\langle\HH^{(-)}(\rr_1,\omega')\otimes\EE^{(+)}(\rr_2,\omega'')\right\rangle&=\sum_{\kk,\sigma,n}\bra{n_{\kk,\sigma}}\HH^{(-)}(\rr_1,\omega')\otimes\EE^{(+)}(\rr_2,\omega'')\ket{n_{\kk,\sigma}}
\nonumber \\ &=\sum_{n}\frac{V}{(2\pi)^3c^3}\int d\Omega \sum_{\sigma}\int d\omega \omega^2 \bra{n_{\kk,\sigma}}\HH^{(-)}(\rr_1,\omega')\otimes\EE^{(+)}(\rr_2,\omega'')\ket{n_{\kk,\sigma}}. \nonumber
\end{align}
Then, using Eqs.~(\ref{e.27}), we can transform this expression into
\begin{align}
&\left\langle\HH^{(-)}(\rr_1,\omega')\otimes\EE^{(+)}(\rr_2,\omega'')\right\rangle\nonumber \\ &=-\frac{1}{ik}\nabla_1\times\left\{\sum_{n}\frac{V}{(2\pi)^3c^3}\int d\Omega \sum_{\sigma}\int d\omega \omega^2 \bra{n_{\kk,\sigma}}\EE^{(-)}(\rr_1,\omega')\otimes\EE^{(+)}(\rr_2,\omega'')\ket{n_{\kk,\sigma}}\right\}, \nonumber
\end{align}
and from here,
\begin{align}\label{e.30}
\left\langle\HH^{(-)}(\rr_1,\omega)\otimes\EE^{(+)}(\rr_2,\omega')\right\rangle=-4\pi\hbar\delta(\omega-\omega')\,\frac{1}{ik}\mbox{Im}\left\{\nabla_1\times \mathcal{G}(\rr_1-\rr_2,\omega)\right\}\,n(\omega).
\end{align}
Likewise, for antinormal ordering, we have
\begin{align}
\left\langle\HH^{(+)}(\rr_1,\omega')\otimes\EE^{(-)}(\rr_2,\omega'')\right\rangle&=\sum_{\kk,\sigma,n}\bra{n_{\kk,\sigma}}\HH^{(+)}(\rr_1,\omega')\otimes\EE^{(-)}(\rr_2,\omega'')\ket{n_{\kk,\sigma}}
\nonumber \\ &=\sum_{n}\frac{V}{(2\pi)^3c^3}\int d\Omega \sum_{\sigma}\int d\omega \omega^2 \bra{n_{\kk,\sigma}}\HH^{(+)}(\rr_1,\omega')\otimes\EE^{(-)}(\rr_2,\omega'')\ket{n_{\kk,\sigma}}, \nonumber
\end{align}
which, following the same procedure as in the previous sections and using Eqs.~(\ref{e.27}), becomes
\begin{align}\label{e.31}
\left\langle\HH^{(+)}(\rr_1,\omega)\otimes\EE^{(-)}(\rr_2,\omega')\right\rangle=4\pi\hbar\delta(\omega-\omega')\,\frac{1}{ik}\mbox{Im}\left\{\nabla_1\times \mathcal{G}(\rr_1-\rr_2,\omega)\right\}\,[n(\omega)+1].
\end{align}

Finally, using Eqs.~(\ref{e.28})-(\ref{e.31}), we can write the symmetrized correlations of the electric-magnetic and the magnetic-electric fluctuations as
\begin{align}
\frac1{2}\left\langle\EE^{(-)}(\rr_1,\omega)\otimes\HH^{(+)}(\rr_2,\omega')+\HH^{(+)}(\rr_2,\omega)\otimes\EE^{(-)}(\rr_1,\omega')\right\rangle=4\pi\hbar\delta(\omega-\omega')\,\frac{1}{ik}\mbox{Im}\left\{\nabla_1\times \mathcal{G}(\rr_1-\rr_2,\omega)\right\}\,[n(\omega)+\frac1{2}], \nonumber
\end{align}
\begin{align}\nonumber
\frac1{2}\left\langle\HH^{(-)}(\rr_1,\omega)\otimes\EE^{(+)}(\rr_2,\omega')+\EE^{(+)}(\rr_2,\omega)\otimes\HH^{(-)}(\rr_1,\omega')\right\rangle=-4\pi\hbar\delta(\omega-\omega')\,\frac{1}{ik}\mbox{Im}\left\{\nabla_1\times \mathcal{G}(\rr_1-\rr_2,\omega)\right\}\,[n(\omega)+\frac1{2}].
\end{align}

\section{Derivation of Eq. (4)}

We follow the notation and definitions introduced in Sec.\ II, and we refer to Fig.\ 1 there for the system under consideration. The net power absorbed by particle 1 can be written as [see Eqs.\ (1) and (2)]
\begin{align}
\mathcal{P}_{1}=\mathcal{P}^{\rm field}_{1}+\mathcal{P}^{\rm dip}_{1} &= \int_{-\infty}^{\infty}\frac{d\omega d\omega'}{(2\pi)^2} e^{-i(\omega-\omega')t} i\omega' \left\langle E_1^{+}(\omega')\left[i\alpha^+(\omega')-(2k^3/3)\left|\alpha(\omega')\right|^2\right]E_1(\omega)\right\rangle \label{r.1} \\ & + \int_{-\infty}^{\infty}\frac{d\omega d\omega'}{(2\pi)^2} e^{-i(\omega-\omega')t}\left[i\w'\left\langle p_1^{+}(\omega')\ggg_{12}(\omega)p_2(\omega)\right\rangle-\frac{2\omega^4}{3c^3} \left\langle p_1^{+}(\omega')p_1(\omega)\right\rangle\right]. \nonumber
\end{align}
Now, using the expressions for the self-consistent fields and dipoles (see Sec.\ II)
\begin{align}
E_1=&D^{-1}\left[E_1^{\rm fl}+\ggg_{12}\al_2E^{\rm fl}_2\right], \nonumber \\
p_1=&D_1^{-1}\left[p_1^{\rm fl}+\al_1\ggg_{12}p^{\rm fl}_2\right], \nonumber \\
p_2=&D_2^{-1}\left[p_2^{\rm fl}+\al_2\ggg_{21}p^{\rm fl}_1\right], \nonumber
\end{align}
where the denominators $D=\II-\ggg_{12}\al_2\ggg_{21}\al_1$, $D_1=\II-\al_1\ggg_{12}\al_2\ggg_{21}$, and $D_2=\II-\al_2\ggg_{21}\al_1\ggg_{12}$ describe multiple scattering, Eq.~(\ref{r.1}) becomes
\begin{align}
\mathcal{P}_{1}=&\int_{-\infty}^{\infty}\frac{d\omega d\omega'}{(2\pi)^2} e^{-i(\omega-\omega')t}\,\nonumber\\
&\times\Bigg\{\omega'\left[\left\langle E_1^{\rm fl +}\left(D^{-1}\right)^+\left[i\alpha^+-(2k^3/3)\left|\alpha\right|^2\right]D^{-1}E_1^{\rm fl}\right\rangle + \left\langle E_1^{\rm fl +} \left(D^{-1}\right)^+\left[i\alpha^+-(2k^3/3)\left|\alpha\right|^2\right]D^{-1}\ggg_{12}\al_2E_2^{\rm fl} \right\rangle \right. \nonumber \\ &\left. \;\;\;\;\;\;\;\;\; +\left\langle E_2^{\rm fl+}\al_2^{+} \ggg_{12}^{+}\left(D^{-1}\right)^+\left[i\alpha^+-(2k^3/3)\left|\alpha\right|^2\right]D^{-1}E_1^{\rm fl}\right\rangle \right. \nonumber \\ &\left.\;\;\;\;\;\;\;\;\;+ \left\langle E_2^{\rm fl +} \al_2^{+} \ggg_{12}^{+}\left(D^{-1}\right)^+\left[i\alpha^+-(2k^3/3)\left|\alpha\right|^2\right]D^{-1}\ggg_{12}\al_2E_2^{\rm fl}\right\rangle\right]\nonumber \\ & \;\;\;\;\;\; + \left[i\w'\left\langle p_1^{\rm fl +} \left(D_1^{-1}\right)^+\ggg_{12}D_2^{-1}\al_2\ggg_{21}p_1^{\rm fl}\right\rangle  + i\w'\left\langle p_2^{\rm fl +}\ggg_{12}^{+}\al_1^{+}\left(D_1^{-1}\right)^+\ggg_{12}D_2^{-1}p_2^{\rm fl} \right \rangle\right. \nonumber \\ & \left. \;\;\;\;\;\;\;\;\;\;\;\; -\frac{2\w^4}{3c^3}\left\langle p_1^{\rm fl+}\left|D_1\right|^{-2}p_1^{\rm fl} \right \rangle -\frac{2\w^4}{3c^3}\left\langle p_2^{\rm fl +}\ggg_{12}^+\al_1^+\left|D_1\right|^{-2}\al_1\ggg_{12}p_2^{\rm fl} \right \rangle\right]\Bigg\} \nonumber\\
=&\sum_{j=1}^8 I_j. \nonumber
\end{align}
We then carry out the matrix multiplications in this expression and apply the FDT. After some algebra, we find
\begin{align}
I_1=\frac{\hbar}{\pi}\int_{0}^{\infty} d\omega \frac{4}{3} \w k^3 \sum_{\nu=E,M}\chipn \left[2\left|\frac{\un}{s}\right|^2+2\left|\frac{\wn}{s}\right|^2+ \left|\frac{1}{\tn}\right|^2\right]\left(n_0+\frac1{2}\right), \nonumber
\end{align}
\begin{align}
I_2+I_3={}&\frac{\hbar}{\pi}\int_{0}^{\infty} d\omega 8\w \sum_{\nu=E,M}\chipn\left[\R\left\{\asn\Ad\left|\frac{\un}{s}\right|^2-\dn\asn\Cd\frac{\un^{*}\wn}{\left|s\right|^2}-\dn\asnp\Cd\frac{\un\wn^{*}}{\left|s\right|^2}+\asnp\Ad\left|\frac{\wn}{s}\right|^2\right\}\I\left\{\Ad\right\}\right.\nonumber\\
&\left.-\I\left\{\asn\Cd\left|\frac{\wn}{s}\right|^2-\dn\asn\Ad\frac{\un\wn^{*}}{\left|s\right|^2}-\dn\asnp\Ad\frac{\un^{*}\wn}{\left|s\right|^2}+\asnp\Ad\left|\frac{\un}{s}\right|^2\right\}\R\left\{\Cd\right\}+\frac1{2}\R\left\{\asn\Bd\left|\frac1{\tn}\right|^2\right\}\I\left\{\Bd\right\}\right]\nonumber \\ &\times\left(n_0+\frac1{2}\right), \nonumber
\end{align}
\begin{align}
I_4=\frac{\hbar}{\pi}\int_{0}^{\infty} d\omega \frac{4}{3} \w k^3 \sum_{\nu=E,M}\chipn\left[2\left|\asn\right|^2\left|\Ad\frac{\un}{s}-\dn\Cd\frac{\wn}{s}\right|^2+2\left|\asnp\right|^2\left|\Ad\frac{\wn}{s}-\dn\Cd\frac{\un}{s}\right|^2+\left|\asn\right|^2\left|\frac{\Bd}{\tn}\right|^2\right]\left(n_0+\frac1{2}\right), \nonumber
\end{align}
\begin{align}
I_5={}&\frac{\hbar}{\pi}\int_{0}^{\infty} d\omega 4 \w \sum_{\nu=E,M}\chipn \I\left\{\frac{\asn}{|s|^2}\left[-\Add\left|\un\right|^2+\dn\Ad\Cd\un^*\wn -\dn\Ad\Cd\un\wn^*+\Add\left|\wn\right|^2\right]\right.\nonumber \\ &\left.+\frac{\asnp}{|s|^2}\left[\Cdd\left|\un\right|^2-\dn\Ad\Cd\un^*\wn+\dn\Ad\Cd\un\wn^*-\Cdd\left|\wn\right|^2\right]-\frac1{2}\asn\Bdd\left|\frac1{\tn}\right|^2 \right. \nonumber \\& - \left.\frac{\wn^*}{|s|^2}\left[\asn\left(\aep\ams-\aes\amp\right)-\asnp\left(\aep\aes-\amp\ams\right)\right]\left(\Ad\Cddd-\Addd\Cd\right)\right\}\left(n_1+\frac1{2}\right), \nonumber
\end{align}
\begin{align}
I_6={}&\frac{\hbar}{\pi}\int_{0}^{\infty} d\omega 4 \w \sum_{\nu=E,M}\chipn \I\left\{\frac{\asn}{|s|^2}\left[\left|\Ad\un\right|^2+\dn\Ad\Cda\un^*\wn -\dn\Ad\Cda\un\wn^*-\left|\Ad\wn\right|^2\right]\right.\nonumber \\ &\left.+\frac{\asnp}{|s|^2}\left[\left|\Cd\unp\right|^2-\dn\Ada\Cd\unp^*\wnp+\dn\Ada\Cd\unp\wnp^*-\left|\Cd\wnp\right|^2\right]+\frac1{2}\asn\left|\frac{\Bd}{\tn}\right|^2 \right. \nonumber \\& + \left.\frac{\wn}{|s|^2}\apn\left(\aep\ams-\aes\amp\right)^*\left(\Ad\Cddda-\Ada\left|\Ad\right|^2\Cda\right)
\right. \nonumber \\& + \left.\frac{\wnp}{|s|^2}\apnp\left(\aep\aes-\amp\ams\right)^*\left(\Ada\left|\Cd\right|^2\Cda-\Addda\Cd\right)
\right\}\left(n_2+\frac1{2}\right), \nonumber
\end{align}
\begin{align}
I_7=-\frac{\hbar}{\pi}\int_{0}^{\infty} d\omega \frac{4}{3} \w k^3 \sum_{\nu=E,M}\chipn \left[2\left|\frac{\un}{s}\right|^2+2\left|\frac{\wn}{s}\right|^2+ \left|\frac{1}{\tn}\right|^2\right]\left(n_1+\frac1{2}\right), \nonumber
\end{align}
\begin{align}
I_8=-\frac{\hbar}{\pi}\int_{0}^{\infty} d\omega \frac{8}{3} \w k^3 \sum_{\nu=E,M}\chisn \left[\left|\apn\right|^2\left|\Ad\frac{\un}{s}-\Cd\frac{\wn}{s}\right|^2  +\left|\apnp\right|^2\left|\Ad\frac{\wnp}{s}+\Cd\frac{\unp}{s}\right|^2+\frac1{2}\left|\frac{\apn \Bd}{\tn}\right|^2\right]\left(n_2+\frac1{2}\right), \nonumber
\end{align}
where the definition of the different variables is the same as in Sec.\ II. In this derivation, we have used the identity
\begin{align}
\I\left\{xy^2\right\}=2\R\left\{xy\right\}\I\left\{y\right\}+\left|y\right|^2\I\left\{x\right\}=2\I\left\{xy\right\}\R\left\{y\right\}-\left|y\right|^2\I\left\{x\right\}. \nonumber
\end{align}
Furthermore, the integral over $\w'$ has been done using the Dirac $\delta$ function of the FDT. After straightforward, lengthy algebraic manipulations, we can rearrange the above terms as
\begin{align}
\mathcal{P}_{1}= H_{01} \left[n_0\left(\w\right)-n_1\left(\w\right)\right]+H_{02}\left[n_0\left(\w\right)-n_2\left(\w\right)\right]+H_{21}\left[n_2\left(\w\right)-n_1\left(\w\right)\right], \label{P1_EPAPS}
\end{align}
where
\begin{align}
H_{01}={}&\frac{\hbar}{\pi} \sum_{\nu=E,M}\int_{0}^{\infty} d\omega 4\w \chipn \left\{\frac{k^3}{3}\left[2\left|\frac{\un}{s}\right|^2+2\left|\frac{\wn}{s}\right|^2+ \left|\frac{1}{\tn}\right|^2\right]-\I\left\{\asn\right\}\left|\Ad\frac{\un}{s}-\dn\Cd\frac{\wn}{s}\right|^2\right. \nonumber \\ \nonumber &\left.-\I\left\{\asnp\right\}\left|\Ad\frac{\wn}{s}-\dn\Cd\frac{\un}{s}\right|^2  +\I\left\{\asn\Add\right\}\left|\frac{\un}{s}\right|^2-\I\left\{\asn\Cdd\right\}\left|\frac{\wn}{s}\right|^2+2\dn\R\left\{\asn\Ad\Cd\right\}\I\left\{\frac{\un\wn^{*}}{|s|^2}\right\}
\right. \\ \nonumber &\left.+\I\left\{\asnp\Add\right\}\left|\frac{\wn}{s}\right|^2-\I\left\{\asnp\Cdd\right\}\left|\frac{\un}{s}\right|^2-2\dn\R\left\{\asnp\Ad\Cd\right\}\I\left\{\frac{\un\wn^{*}}{|s|^2}\right\}+\R\left\{\asn \Bd\right\}\I\left\{\Bd\right\}\left|\frac{1}{\tn}\right|^2 \right\}, \nonumber
\end{align}
\begin{align}
H_{02}={}&\frac{\hbar}{\pi} \sum_{\nu=E,M}\int_{0}^{\infty} d\omega \w\frac{8k^3}{3}\chipn \left\{\left|\asn\right|^2\left|\Ad\frac{\un}{s}-\dn\Cd\frac{\wn}{s}\right|^2 + \left|\asnp\right|^2\left|\Ad\frac{\wn}{s}-\dn\Cd\frac{\un}{s}\right|^2+\frac1{2}\left|\asn\frac{\Bd}{\tn}\right|\right\}, \nonumber
\end{align}
\begin{align}
H_{21}={}&\frac{\hbar}{\pi} \sum_{\nu=E,M}\int_{0}^{\infty} d\omega 4\w\chipn \left\{\I\left\{\asn\right\}\left|\Ad\frac{\un}{s}-\dn\Cd\frac{\wn}{s}\right|^2 + \I\left\{\asnp\right\}\left|\Ad\frac{\wn}{s}-\dn\Cd\frac{\un}{s}\right|^2+\frac{1}{2}\I\left\{\asn\right\}\left|\frac{\Bd}{\tn}\right|^2\right\}. \nonumber
\end{align}
Finally, Eq.\ (4) can be readily obtained by reorganizing these expressions.

\begin{figure}
\begin{center}
\includegraphics[width=130mm,angle=0,clip]{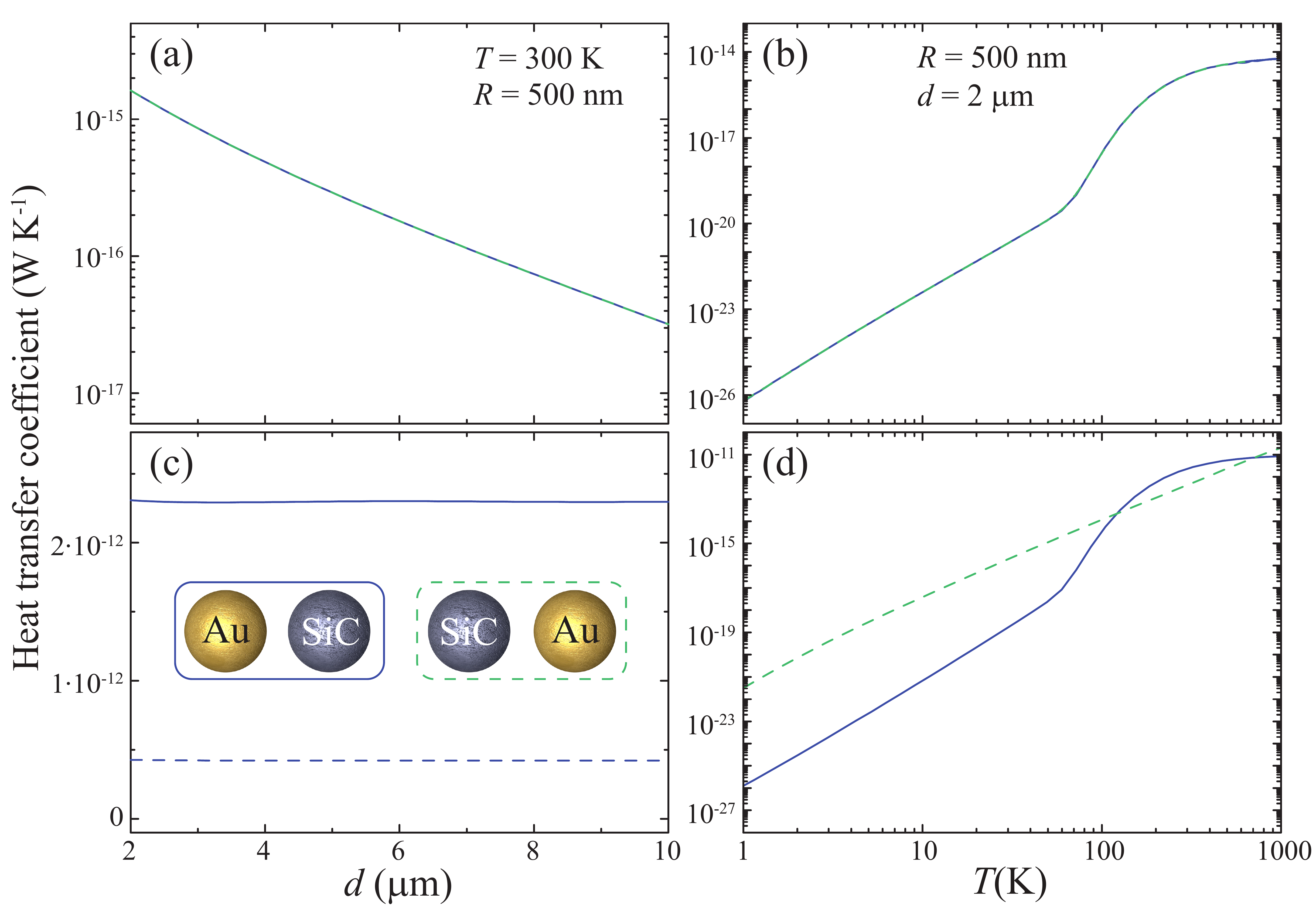}
\caption{{\bf (a,b)} Heat transfer coefficient (HTC) as a function of particle distance $d$ and temperature $T$ for a gold-SiC dimer (solid curves) and a SiC-god dimer (broken curves). The HTC from the right particle (2) to the left particle (1) is defined as $\mathcal{P}_1/\delta T$ [see Eq.\ (\ref{P1_EPAPS})] with $T_1=T_0=T$ and $T_2=T+\delta T$ [see insets in (c) for sketches of the dimers]. {\bf (c,d)} Power emanating from the right particle 
in the two dimer orientations under consideration.} \label{Fig8}
\end{center}
\end{figure}

\section{Symmetric and asymmetric heat transfers in inhomogeneous dimers}

We show in Fig.\ \ref{Fig8}(a,b) the different heat transfer coefficients (HTCs, see Sec.\ III) exhibited by an inhomogeneous gold-SiC dimer when the hotter particle is either SiC (solid curves) or gold (dashed curves) and the transfer is for heat absorbed by the remaining gold or SiC particle, respectively. The power transfer to the cooler particle is nearly independent on whether the hotter particle is SiC or Au. This symmetry upon permutation of particle indices ($1\leftrightarrow2$) is complete if we neglect multiple scattering between the particles [i.e., by setting $u_\nu=1$ and $w_\nu=0$ in Eq.\ (4), since only $\Gamma^\nu_{2\perp}$ and $\Gamma^\nu_{2\parallel}$ terms contribute to the HTC], and indeed multiple scattering can be neglected in the clusters under consideration, as shown in Figs. 2-4.

In contrast, the total heat lost by the hotter particle is strongly dependent on whether this is gold or SiC [see Fig.\ \ref{Fig8}(c,d)]. Part of this heat is absorbed by the cooler particle, but the rest is radiated into the surrounding vacuum. At high temperatures above $\sim100\,$K, a hotter SiC particle produces larger radiation rates [Fig.\ \ref{Fig8}(d)], rather independent of particle distance [Fig.\ \ref{Fig8}(c)]. However, hotter gold is more capable of radiating at smaller temperatures [cf. vertical scales in Fig.\ \ref{Fig8}(b,d)].

\begin{figure}
\begin{center}
\includegraphics[width=160mm,angle=0,clip]{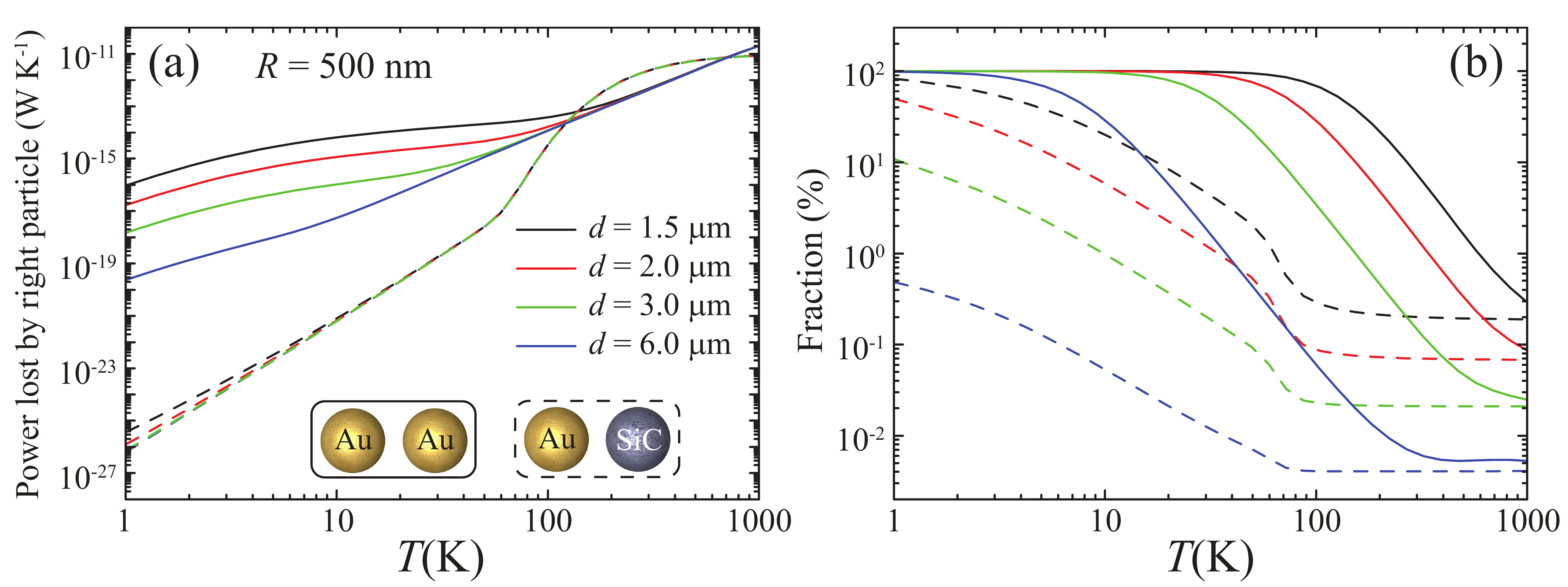}
\caption{{\bf (a)} Heat power lost by the right particle as a function of temperature for different particle separations. The right particle is slightly hotter than both the vacuum and the left particle, which are prepared at the same temperature, and the rate is normalized to the small temperature difference of the right particle (i.e., we represent $-\mathcal{P}_{2}/\delta T$ for $T_0=T_1=T$ and $T_2=T+\delta T$). {\bf (b)} Fraction of the power lost by the right particle that ends up being absorbed by the left particle.} \label{Fig9}
\end{center}
\end{figure}

\section{Total rate of heat loss and fraction of power exchange}

Heat exchange with the surrounding vacuum plays a leading role in the energy balance during radiative thermalization of a particle dimer, as shown in Fig.\ \ref{Fig9}. We plot in Fig.\ \ref{Fig9}(a) the power lost by the right particle when it is placed at a slightly higher temperature with respect to both the vacuum and the left particle (i.e., we represent $-\mathcal{P}_{2}/\delta T$ for $T_0=T_1=T$ and $T_2=T+\delta T$). At high temperatures, this power is rather insensitive to the presence of the neighboring particle and no significant dependence on particle separation $d$ is observed. This happens above $T\sim50\,$K in the inhomogeneous gold-SiC dimer and above $T\sim300\,$K in the homogeneous gold-gold dimer. However, the accompanying left particle has a strong influence on the power loss of the right particle at lower temperatures, specially in the homogeneous dimer. The power loss increases when the particles are placed closer together. We show next that this is partially explained by the effect of additional absorption by the left\ particle.

Figure\ \ref{Fig9}(b) shows the fraction of the power lost by the right particle that is absorbed by the left particle. This fraction drops to small values at large temperatures, but it eventually approaches 100\% at lower temperatures. This behavior is consistent with the distance dependence of the power lost by the right particle. Nearly full radiative heat transfer between two neighboring particles with negligible radiation into the surrounded vacuum is thus possible at sufficiently low temperatures. Specifically, in the gold-gold dimer this regime is already achieved at $\sim100\,$K for particles of radius $R=500\,$nm and a surface-to-surface separation of one radius (i.e., $d=1.5\,\mu$m). We observe that the temperature below which nearly 100\% transfer between the dimer particles takes place decreases with increasing separation.

%\bibliographystyle{apsrev}
%\bibliography{../../../bibtex/refs}

\end{document}